\documentclass[useAMS,usenatbib]{mnras}
\usepackage{journals}
\usepackage{graphicx}
\usepackage{color}
\usepackage{times}
\usepackage{hyperref}
\usepackage{amsmath}

\title[Supernovae and their host galaxies -- III]{Supernovae and their host galaxies -- III.
The impact of bars and bulges on the radial distribution of supernovae in disc galaxies}
\author[A.~A.~Hakobyan~et~al.]{A.~A.~Hakobyan,$^{1}$\thanks{E-mail:
hakobyan@bao.sci.am}
A.~G.~Karapetyan,$^{1}$
L.~V.~Barkhudaryan,$^{1}$
G.~A.~Mamon,$^{2}$
\newauthor
D.~Kunth,$^{2}$
A.~R.~Petrosian,$^{1}$
V.~Adibekyan,$^{3}$
L.~S.~Aramyan$^{1}$
and M.~Turatto$^{4}$
\\
$^{1}$Byurakan Astrophysical Observatory, 0213 Byurakan, Aragatsotn province, Armenia\\
$^{2}$Institut d'Astrophysique de Paris (UMR 7095: CNRS \& UPMC), 98 bis Bd Arago, F-75014 Paris, France\\
$^{3}$Centro de Astrof\'{i}sica da Universidade do Porto, Rua das Estrelas, P-4150-762 Porto, Portugal\\
$^{4}$INAF -- Osservatorio Astronomico di Padova, Vicolo dell'Osservatorio 5, I-35122 Padova, Italy}
\begin{document}

\date{Accepted 2015 December 2. Received 2015 November 28; in original form 2015 October 16}

\pagerange{\pageref{firstpage}--\pageref{lastpage}} \pubyear{2016}

\maketitle

\label{firstpage}

\begin{abstract}
  We present an analysis of the impact of bars and bulges on the radial distributions of
  the different types of supernovae (SNe) in the stellar discs of host galaxies with various morphologies.
  We use a well-defined sample of 500 nearby (${\leq {\rm 100~Mpc}}$) SNe and their
  low-inclined ($i \leq 60^\circ$) and morphologically non-disturbed
  S0--Sm host galaxies from the Sloan Digital Sky Survey.
  We find that in Sa--Sm galaxies, all core-collapse (CC) and vast majority of SNe Ia belong to the disc,
  rather than the bulge component.
  The radial distribution of SNe Ia in S0--S0/a galaxies is
  inconsistent with their distribution in Sa--Sm hosts, which is
  probably due to the contribution of the outer bulge SNe Ia
  in S0--S0/a galaxies.
  In Sa--Sbc galaxies, the radial distribution of CC SNe in barred hosts is inconsistent with that
  in unbarred ones, while the distributions of SNe Ia are not significantly different.
  At the same time, the radial distributions of both types of SNe in Sc--Sm galaxies are not affected by bars.
  We propose that the additional mechanism shaping the distributions of Type Ia and CC SNe can be explained within the framework of substantial suppression of massive star formation in the radial range
  swept by strong bars, particularly in early-type spirals.
  The radial distribution of CC SNe in unbarred Sa--Sbc galaxies is more centrally peaked
  and inconsistent with that in unbarred Sc--Sm hosts, while the distribution
  of SNe Ia in unbarred galaxies is not affected by host morphology.
  These results can be explained by the distinct distributions
  of massive stars in the discs of early- and late-type spirals.
\end{abstract}


\section{Introduction}
\label{intro}

In the realm of disc galaxies, bars are a common feature
observed in approximately 40 per cent of nearby S0--Sm galaxies
(e.g. \citealt*[][]{2011A&A...532A..75D}).
In the central regions of these disc galaxies,
about one-third have strong barred structure,
which generally affects both the motions of stars and interstellar gas
and can affect spiral arms as well
\citep[for a comprehensive review see][]{2004ARA&A..42..603K}.
In addition, the relative sizes of bars and bulges seems to be correlated
\citep[e.g.][]{2009chas.book..159G},
thus indicating that the growth of bars and bulges are somehow connected.

In spiral galaxies, most of the massive star formation occurs in the discs
where bars, if present, have a strong impact on the radial distribution of
star formation, particularly in early Hubble types.
\citet*[][]{2009A&A...501..207J} used H$\alpha$ and $R$-band imaging to
determine the distributions of young and old stellar populations in several hundreds of
nearby S0/a--Im field galaxies.
They identified a clear effect of bars on the pattern of massive star formation as
a function of radius within discs.
This effect results in a strongly enhanced H$\alpha$ emission,
and moderately enhanced $R$-band emission in both the central regions and at the bar-end radius
of galaxies \citep[see also][]{2015MNRAS.450.3503J}.
The authors noted that this effect seems to be stronger in
galaxies classified as barred Sb or Sbc,
where the overall distributions of star formation markedly different
from that in their unbarred counterparts.

In this context, we are investigating the possible impact of
stellar bars and bulges on the radial distributions of the different types of supernovae (SNe)
in S0--Sm host galaxies.
SNe are generally divided into two categories according to their progenitors:
core-collapse (CC) and Type Ia SNe.
CC SNe result from massive young stars that undergo CC
\citep[e.g.][]{2003LNP...598...21T,2009ARA&A..47...63S,
2012MNRAS.424.1372A},\footnote{{\footnotesize CC SNe
are observationally classified in three major classes, according to
the strength of lines in optical spectra \citep[e.g.][]{1997ARA&A..35..309F}:
Type II SNe show hydrogen lines in
their spectra, including the IIn (dominated by emission lines with narrow components)
and IIb (transitional objects with observed properties closer to SNe II at early times,
then metamorphosing to SNe Ib) subclasses;
Type Ib SNe show helium but not hydrogen, while Type Ic SNe show neither hydrogen nor helium.
All these SNe types arise from young massive progenitors with possible differences in
their masses, metallicities, ages, and fractions of binary stellar systems
\citep[e.g.][]{2011MNRAS.412.1522S}.}}
while Type Ia SNe are the end point in the evolution of binary stars when
an older white dwarf (WD) accretes material from its companion,
causing the WD mass to exceed the Chandrasekhar limit,
or a double WD system loses angular momentum due to gravitational wave
emission, leading to coalescence and explosion
\citep*[for a comprehensive review,
including the scenarios where the WD explodes at masses both above
and below the Chandrasekhar limit, see][]{2014ARA&A..52..107M}.
The distribution of Type Ia SNe traces the distribution of
$R$-band continuum emission/stellar mass in host disc galaxies,
while the distribution of CC SNe is strongly related to the distribution of
H$\alpha$ emission/star formation
\citep[for the most recent review, see][]{2015PASA...32...19A}.
Therefore, it is expected that the above mentioned bar effect and
a possible contribution from the old bulge populations in disc galaxies would leave their
`fingerprints' on the radial distributions of Type Ia and CC SNe.
\defcitealias{2009A&A...508.1259H}{H09}

In our earlier study (\citealp{2009A&A...508.1259H}, hereafter H09),
we attempted to find differences in the radial distributions of SNe
in barred and unbarred spiral host galaxies
\citep[see also][]{2005AJ....129.1369P,2008Ap.....51...69H,2013Ap&SS.347..365N}.
No significant differences were found, most probably due to
small number statistics, inhomogeneous data sets of SNe and their hosts,
and unsatisfying considerations for the relations between the bar lengths, bulge sizes,
and morphological types of SNe host galaxies.
On the other hand, \citet*[][]{1997ApJ...483L..29W}
attempted to estimate the contribution from bulge components of
spiral host galaxies to the entire radial distribution of SNe Ia.
They noted that the stellar bulges in spirals are not efficient producers of Type Ia SNe
\citep[see also][]{2015MNRAS.448..732A}.
But again, the above mentioned impact of bars on the radial distributions of
young and old stellar populations in the discs was not considered.

The aim of this article is to address these questions properly through a study of
the radial distributions of Type Ia and CC (Ibc and II) SNe in a well-defined and homogeneous
sample of 500 nearby SNe and
their low-inclined and morphologically non-disturbed S0--Sm galaxies
from the coverage of Sloan Digital Sky Survey-III \citep[SDSS-III;][]{2014ApJS..211...17A}.
\defcitealias{2012A&A...544A..81H}{I}
\defcitealias{2014MNRAS.444.2428H}{II}

In our first paper of this series \citep[][hereafter Paper~I]{2012A&A...544A..81H},
we have created a large and well-defined data base that combines extensive new measurements
and a literature search of 3876 SNe and their 3679 host galaxies located within the sky area covered by
the SDSS Data Release 8 (DR8).
This data base is much larger than previous ones, and provides a homogeneous set of
global parameters of SN hosts, including morphological classifications and measures of
activity classes of nuclei.
Moreover, in Paper~\citetalias{2012A&A...544A..81H}, we analysed and discussed many
selection effects and biases, which usually affect the statistical studies of SNe.
In the second article of the series \citet[][hereafter Paper~II]{2014MNRAS.444.2428H},
we presented an analysis of the relative frequencies of the different SN types in nearby spiral
galaxies with various morphological types and with or without bars.
We used a subsample of spiral host galaxies of 692 SNe in different stages of galaxy--galaxy interaction
and activity classes of nucleus. We proposed that the underlying mechanisms shaping the number ratios
of SNe types could be interpreted within the framework of interaction-induced star formation,
in addition to the known relations between morphologies and stellar populations.
For more details, the reader is referred to Papers~\citetalias{2012A&A...544A..81H} and
\citetalias{2014MNRAS.444.2428H}.

This is the third paper of the series and
the outline is as follows. Section~\ref{sample} introduces
sample selection and reduction.
In Section~\ref{resdiscus}, we give the results and discuss
all the statistical relations.
Our conclusions are summarized in
Section~\ref{concl}.
Throughout this paper, we adopt a cosmological model with
$\Omega_{\rm m}=0.27$, $\Omega_{\rm \Lambda}=0.73$, and a Hubble constant is taken as
$H_0=73 \,\rm km \,s^{-1} \,Mpc^{-1}$ \citep{2007ApJS..170..377S},
both to conform to values used in our data base.

\section{Sample selection and reduction}
\label{sample}

In this study, we compiled our sample by cross-matching the coordinates of
classified Ia, Ibc\footnote{{\footnotesize By SN~Ibc, we
denote stripped-envelope SNe of Types Ib and Ic, as well as mixed Ib/c
whose specific subclassification is uncertain.}}, and II SNe from the
Asiago Supernova Catalogue\footnote{{\footnotesize We use
the updated version of the ASC to include all classified SNe exploded
before 2014 January 1.}}
\citep[ASC;][]{1999A&AS..139..531B}
with the coverage of SDSS DR10 \citep[][]{2014ApJS..211...17A}.
All SNe are required to have coordinates and/or positions (offsets)
with respect to the nuclei of their host galaxies.
We use SDSS DR10 and the techniques presented in Paper~\citetalias{2012A&A...544A..81H} to
identify the SNe host galaxies and classify their morphological types.
Since we are interested in studying the radial distribution of SNe in
stellar discs of galaxies, the morphologies of hosts are restricted to S0--Sm types.

In Paper~\citetalias{2012A&A...544A..81H},
we have shown that the sample of SNe is largely incomplete beyond ${\rm 100~Mpc}$
(see also Paper~\citetalias{2014MNRAS.444.2428H}).
Thus, to avoid biasing the current sample against or in favour of
one of the SN types, we truncate the sample to distances ${\leq {\rm 100~Mpc}}$.

In addition, following the approach described in detail in
Paper~\citetalias{2014MNRAS.444.2428H}, we classify
the morphological disturbances
of the host galaxies from the visible
signs of galaxy--galaxy interactions in the SDSS DR10. We adopted
the following categories for SN host disturbances: normal (hosts without any
visible disturbance in their morphological structure), perturbed (hosts
with visible morphological disturbance, but without long tidal arms,
bridges, or destructed spiral patterns), interacting (hosts with
obvious signs of galaxy--galaxy interaction), merging (hosts with
ongoing merging process), and post-merging/remnant
(single galaxies that exhibit signs of a past interaction, with a strong
or relaxed disturbance).
Here, we make use of this classification in order to
exclude from the present analysis any host galaxy exhibiting
strong disturbances (interacting, merging, and post-merging/remnant).

\begin{table}
  \centering
  \begin{minipage}{83mm}
  \caption{Numbers of SNe at distances $\leq 100~{\rm Mpc}$ and inclinations
           $i\leq 60^\circ$ within host disc galaxies as a function of
           morphological types, split between barred and unbarred.}
  \tabcolsep 3.5pt
  \label{table_SN_morph}
  \begin{tabular}{lrrrrrrrrrrrr}
  \hline
   &\multicolumn{12}{c}{Barred}\\
  \cline{2-13}
   &\multicolumn{1}{c}{S0}&\multicolumn{1}{c}{S0/a}&\multicolumn{1}{c}{Sa}
  &\multicolumn{1}{c}{Sab}&\multicolumn{1}{c}{Sb}&\multicolumn{1}{c}{Sbc}&\multicolumn{1}{c}{Sc}&\multicolumn{1}{c}{Scd}
  &\multicolumn{1}{c}{Sd}&\multicolumn{1}{c}{Sdm}&\multicolumn{1}{c}{Sm}&\multicolumn{1}{r}{All}\\
  \hline
  Ia & 1 & 9 & 5 & 8 & 13 & 13 & 13 & 6 & 12 & 3 & 0 & 83 \\
  Ib & 0 & 0 & 0 & 0 & 1 & 5 & 3 & 0 & 1 & 2 & 1 & 13 \\
  Ib/c & 0 & 0 & 0 & 0 & 0 & 2 & 0 & 0 & 0 & 0 & 0 & 2 \\
  Ic & 0 & 0 & 0 & 0 & 1 & 3 & 5 & 4 & 4 & 0 & 0 & 17 \\
  II & 0 & 0 & 0 & 2 & 15 & 17 & 19 & 8 & 25 & 8 & 2 & 96 \\
  IIb & 0 & 0 & 1 & 0 & 0 & 3 & 2 & 0 & 2 & 0 & 0 & 8 \\
  \hline
  All & 1 & 9 & 6 & 10 & 30 & 43 & 42 & 18 & 44 & 13 & 3 & 219 \\
  \hline
   &\multicolumn{12}{c}{Unbarred}\\
  \cline{2-13}
   &\multicolumn{1}{c}{S0}&\multicolumn{1}{c}{S0/a}&\multicolumn{1}{c}{Sa}
  &\multicolumn{1}{c}{Sab}&\multicolumn{1}{c}{Sb}&\multicolumn{1}{c}{Sbc}&\multicolumn{1}{c}{Sc}&\multicolumn{1}{c}{Scd}
  &\multicolumn{1}{c}{Sd}&\multicolumn{1}{c}{Sdm}&\multicolumn{1}{c}{Sm}&\multicolumn{1}{r}{All}\\
  \hline
  Ia & 7 & 11 & 4 & 4 & 10 & 22 & 26 & 5 & 4 & 0 & 4 & 97 \\
  Ib & 0 & 0 & 0 & 1 & 0 & 7 & 8 & 2 & 0 & 0 & 0 & 18 \\
  Ib/c & 0 & 0 & 0 & 1 & 3 & 0 & 1 & 1 & 1 & 0 & 0 & 7 \\
  Ic & 0 & 0 & 0 & 0 & 1 & 6 & 13 & 2 & 1 & 0 & 0 & 23 \\
  II & 1 & 0 & 0 & 0 & 15 & 22 & 64 & 13 & 6 & 3 & 2 & 126 \\
  IIb & 0 & 0 & 0 & 0 & 0 & 0 & 5 & 3 & 0 & 2 & 0 & 10 \\
  \hline
  All & 8 & 11 & 4 & 6 & 29 & 57 & 117 & 26 & 12 & 5 & 6 & 281 \\
  \hline \\
  \end{tabular}
  \parbox{\hsize}{Among the SNe types, there are only 31 uncertain
           (`:' or `*') and 35 peculiar (`pec') classifications.
           All Type IIn SNe are removed from the sample due to uncertainties in their progenitor nature,
           and often in their classification \citep[e.g.][]{2012MNRAS.424.1372A,2014MNRAS.441.2230H}.}
  \end{minipage}
\end{table}

We measure the geometry of host galaxies using the Graphical Astronomy and Image
Analysis\footnote{{\footnotesize GAIA is available for download as part of
JAC Starlink Release at \texttt{http://starlink.jach.hawaii.edu}.}}
(GAIA) tool according to the approaches
presented in Paper~\citetalias{2012A&A...544A..81H}.\footnote{{\footnotesize The
data base of Paper~\citetalias{2012A&A...544A..81H} is based on the SDSS DR8.
Here, because we added new SNe in the sample, for homogeneity we
re/measure the geometry of all host galaxies based only on DR10.}}
First, we construct $25~{\rm mag~arcsec^{-2}}$ isophotes in the SDSS DR10 $g$-band,
and then we visually fit on to each isophote an elliptical aperture centred at each
galaxy centroid position.
From the fitted elliptical apertures, we derive the major axes ($D_{25}$),
elongations ($a/b$), and position angles (PA) of the major axes of
galaxies. In further analysis, we use the $D_{25}$
corrected for Galactic and host galaxy internal extinction.
We then calculate the inclinations of host galaxies using elongations
and morphological types following the classical Hubble formula.
More details on these procedures are found in Paper~\citetalias{2012A&A...544A..81H}.
Finally, an additional restriction on the inclinations ($i \leq 60^\circ$) of hosts
is required to minimize absorption and projection effects in the discs of
galaxies.\footnote{{\footnotesize The sample with $i > 60^\circ$
is only used in Section~\ref{resdiscus_sub1} for ancillary purposes.}}

After these restrictions, we are left with a sample of
500 SNe within 419 host galaxies.\footnote{{\footnotesize The full data base of
500 individual SNe (SN designation, type, and offset from host galaxy nucleus) and
their 419 host galaxies (galaxy SDSS designation, distance, morphological type, bar,
corrected $g$-band $D_{25}$, $a/b$, PA, and inclination) is available in the
online version of this article.}}
For these host galaxies, we do visual inspection of the combined
SDSS $g$-, $r$-, and $i$-band images, as well as check the different bands separately
to detect any sign of barred structure in the stellar discs.
In Paper~\citetalias{2012A&A...544A..81H}, we demonstrated that given their
superior angular resolution and three-colour representations, the SDSS images,
especially for low-inclined galaxies, offer
a much more reliable and capacious source for bar detection than do the other images,
e.g. plate-based images on which most of
the HyperLeda/NED classifications were performed.
To check the consistency of bar detection in our sample with that of
the SDSS-based EFIGI\footnote{{\footnotesize
Extraction de Formes Id\'{e}alis\'{e}es de Galaxies en Imagerie.}}
catalogue of nearby visually classified 4458 PGC galaxies \citep{2011A&A...532A..74B},
we select a subsample of 149 galaxies that are common to both EFIGI and to our sample of SNe hosts.
Following Paper~\citetalias{2014MNRAS.444.2428H},
we compare the EFIGI \texttt{Bar Length} attribute\footnote{{\footnotesize
This attribute quantifies the presence of a central bar component in the galaxy,
in terms of length relative to the galaxy major axis $D_{25}$.
\texttt{Bar Length} $=$
0 (no visible bar); 1 (barely visible bar feature);
2 (short bar, with a length about one-third of $D_{25}$);
3 (long bar, that extends over about half of $D_{25}$);
4 (very long, prominent bar that extends over more than half of $D_{25}$).}}
with our detection (bar or no bar). When the \texttt{Bar Length} is 0 or 2,
our bar detection is different for only 6 per cent of cases.
When the \texttt{Bar Length} is 3 or 4,
our bar detection completely matches with that of the EFIGI.
However, we do not detect bars in 30 per cent of the cases when \texttt{Bar Length} is 1.
Hence, the EFIGI \texttt{Bar Length} $=$ 1 mainly corresponds to the threshold of our bar detection,
i.e., barely visible bar feature with a length about one-tenth of $D_{25}$.\footnote{{\footnotesize
Later in the study, we will see that this very central region of galaxies
with tiny bars is almost always restricted to the first bin of the radial distribution of SNe,
and therefore cannot statistically bias our results.
}}
Table~\ref{table_SN_morph} displays the distribution of all SNe types
among the various considered morphological types for their
barred and unbarred host galaxies.

\section{Results and discussion}
\label{resdiscus}

We now study the possible influence of bars and bulges on
the SNe distributions through an analysis of the radial distributions of
different types of SNe in discs of host galaxies with various morphological types.
Here, we widely use our morphological classification, bar detection,
and geometry for SNe host galaxies, while for bulge parameters
(e.g. bulge-to-disc mass ratio: B/D)
we only indirectly consider their statistical relations to the Hubble sequence
\citep[e.g.][]{2011A&A...532A..75D}.

\subsection{Galactocentric radius deprojection and normalization}
\label{resdiscus_sub1}

We apply the inclination correction to the projected (observed)
galactocentric radii of SNe as described in \citetalias{2009A&A...508.1259H}.
The reliability of the inclination correction is based on the fundamental assumption
that SNe belong mostly to the disc, rather than the bulge component in
S0--Sm galaxies \cite[e.g.][]{1963PASP...75..123J,1973RA......8..411M,1975A&A....44..267B,
1975PASJ...27..411I,1977MNRAS.178..693V,1981SvAL....7..254T,1987SvA....31...39T,1997Ap.....40..296K,
2000ApJ...542..588I,2013Sci...340..170W}.
This assumption is natural for CC SNe, considering the requirement that
massive young stars are progenitors \cite[e.g.][]{2009ARA&A..47...63S,2012MNRAS.424.1372A}
located in the discs of host galaxies
(e.g. \citealt{2008RMxAA..44..103O}; \citetalias{2009A&A...508.1259H}).
The spatial distribution of SNe Ia, arising from older WD in binary systems
\cite[e.g.][]{2014ARA&A..52..107M},
is more complicated and consists of two interpenetrating components
(e.g. \citealt*{1998ApJ...502..177H}): disc and bulge SNe.

If SNe belong mostly to the bulge component, one would expect that the distributions of
their projected galactocentric distances (normalized to the $g$-band $25^{\rm th}$ magnitude isophotal
semimajor axis; $R_{25}=D_{25}/2$) along major ($U/R_{25}$) and minor ($V/R_{25}$)
axes would be the same both in face-on ($i \leq 60^\circ$) and in edge-on ($i > 60^\circ$) disc galaxies.
The projected $U$ and $V$ galactocentric distances (in arcsec) of an SN are
\begin{equation}
\begin{split}
 U = \Delta \alpha \,\sin {\rm PA} + \Delta \delta \,\cos {\rm PA} \ ,\\
 V = \Delta \alpha \,\cos {\rm PA} - \Delta \delta \,\sin {\rm PA} \ , \\
\end{split}
\label{UVdefs}
\end{equation}
where $\Delta\alpha$ and $\Delta\delta$ are offsets of the SN in equatorial coordinate system,
and PA is position angle of the major axis of the host galaxy.
For more details of these formulae, the reader is referred to \citetalias{2009A&A...508.1259H}.

The $R_{25}$ normalization is important because
the distribution of linear distances (in kpc) is strongly biased by
the greatly different intrinsic sizes of host galaxies
(see fig.~2 in \citetalias{2009A&A...508.1259H}).
In linear scale, there would be a systematic overpopulation of
SNe at small galactocentric distances as this region would be populated by
SNe exploding in all host galaxies, including the smaller ones,
while larger distances would only be populated by the SNe occurring in the larger hosts.

\begin{figure*}
\begin{center}$
\begin{array}{@{\hspace{0mm}}l@{\hspace{0mm}}r@{\hspace{0mm}}}
\includegraphics[width=0.45\hsize]{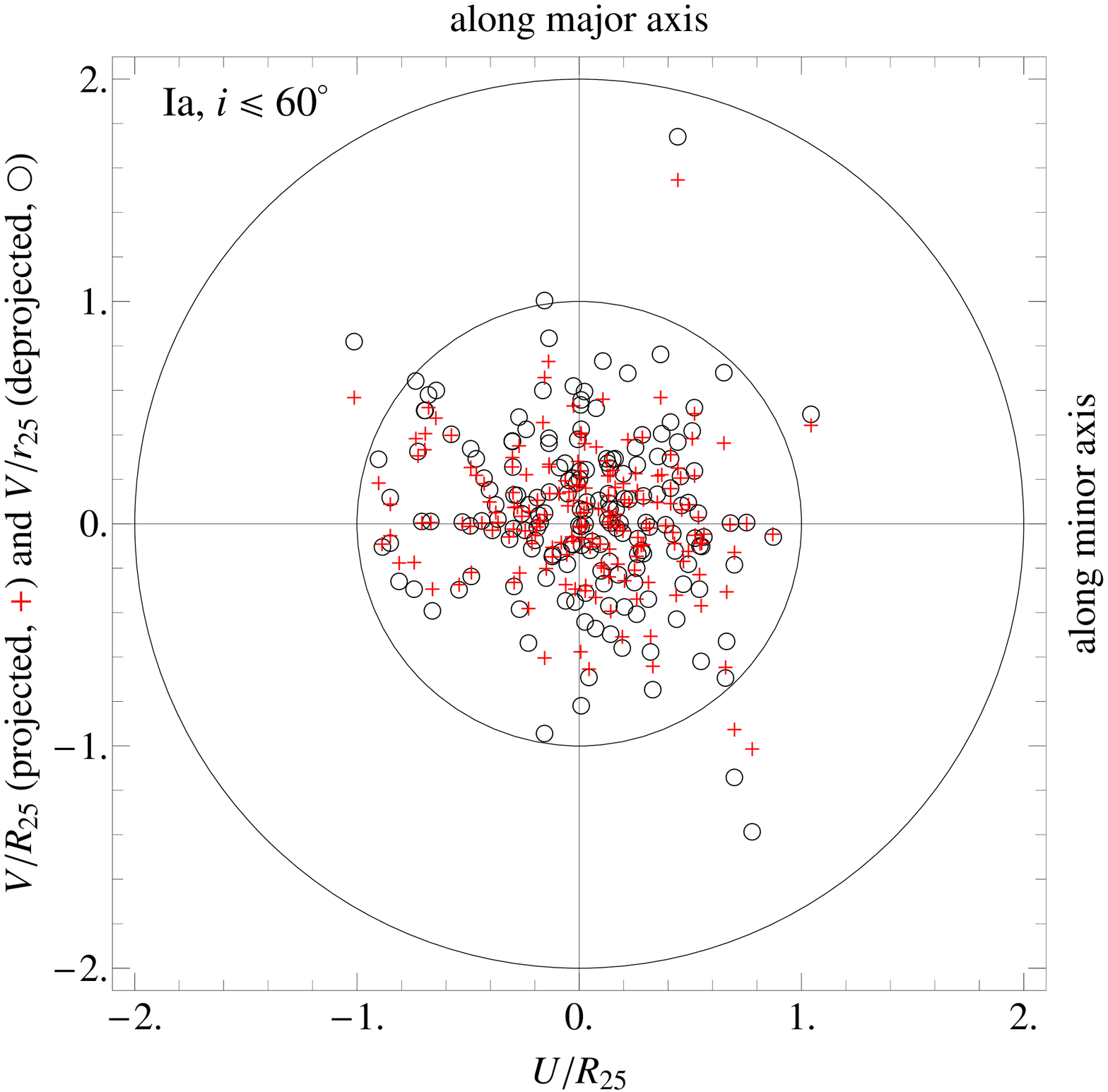} &
\includegraphics[width=0.45\hsize]{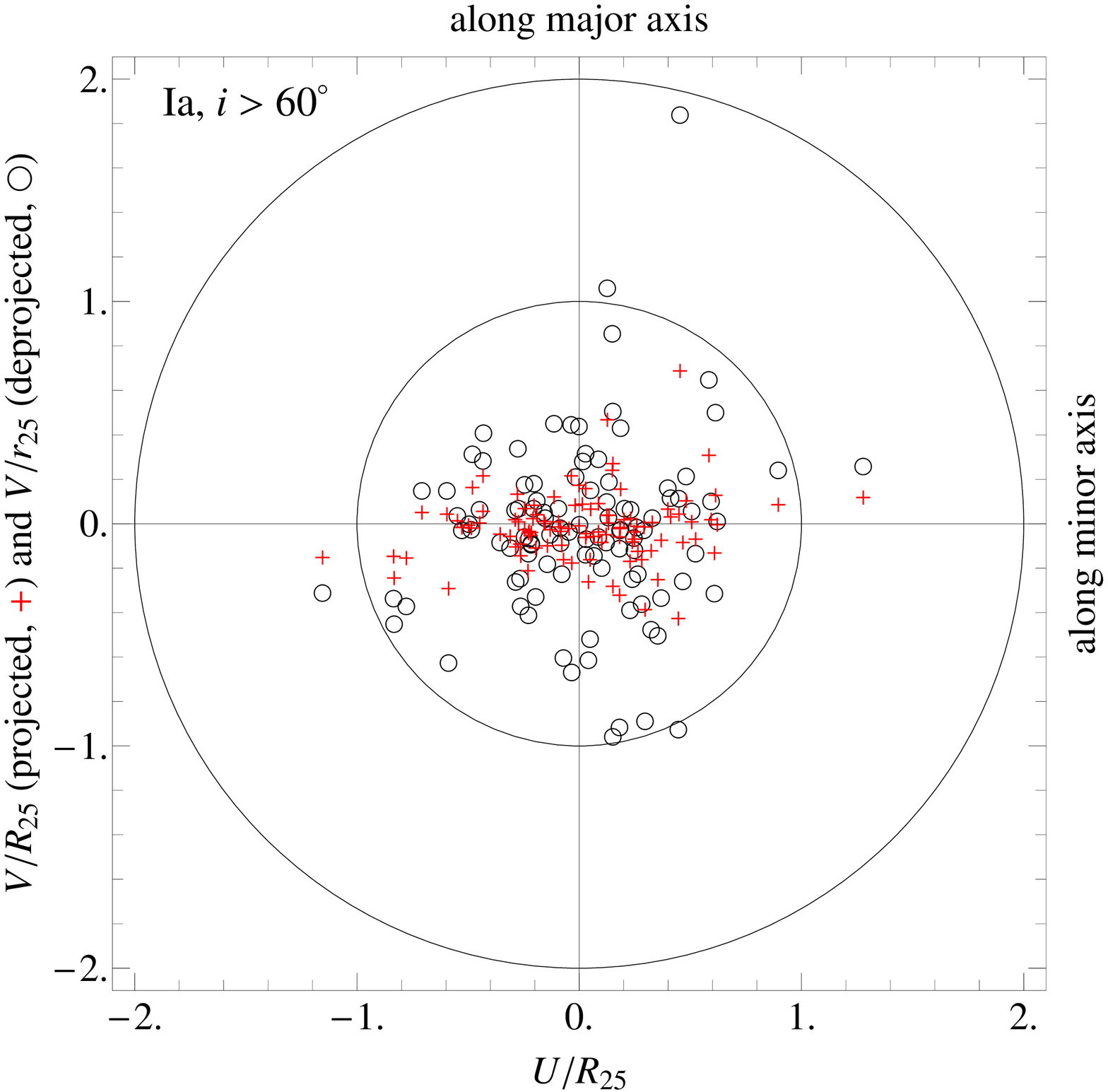}
\end{array}$
\end{center}
\begin{center}$
\begin{array}{@{\hspace{0mm}}r@{\hspace{0mm}}r@{\hspace{5mm}}r@{\hspace{0mm}}r@{\hspace{0mm}}}
\includegraphics[width=0.22\hsize]{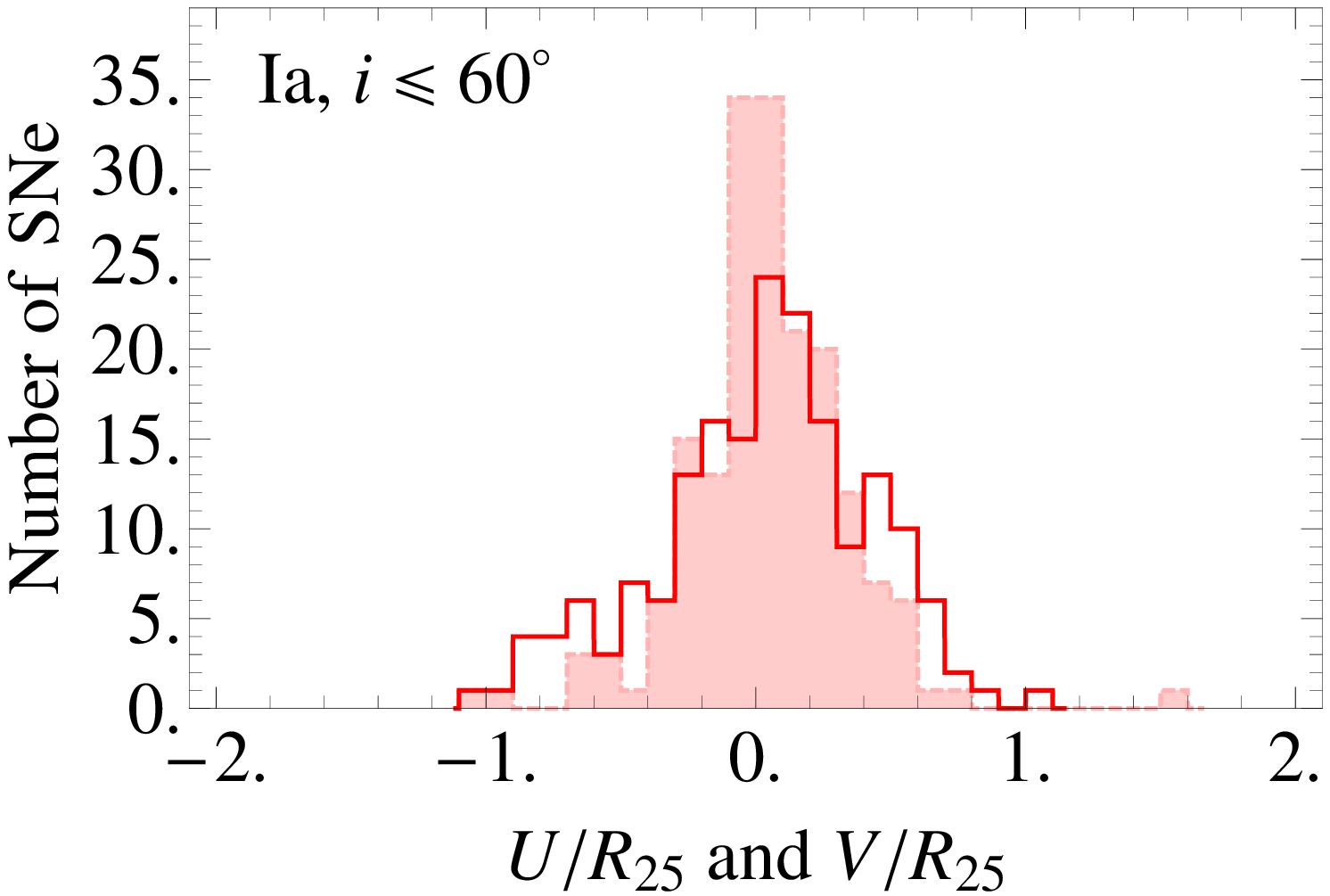} &
\includegraphics[width=0.22\hsize]{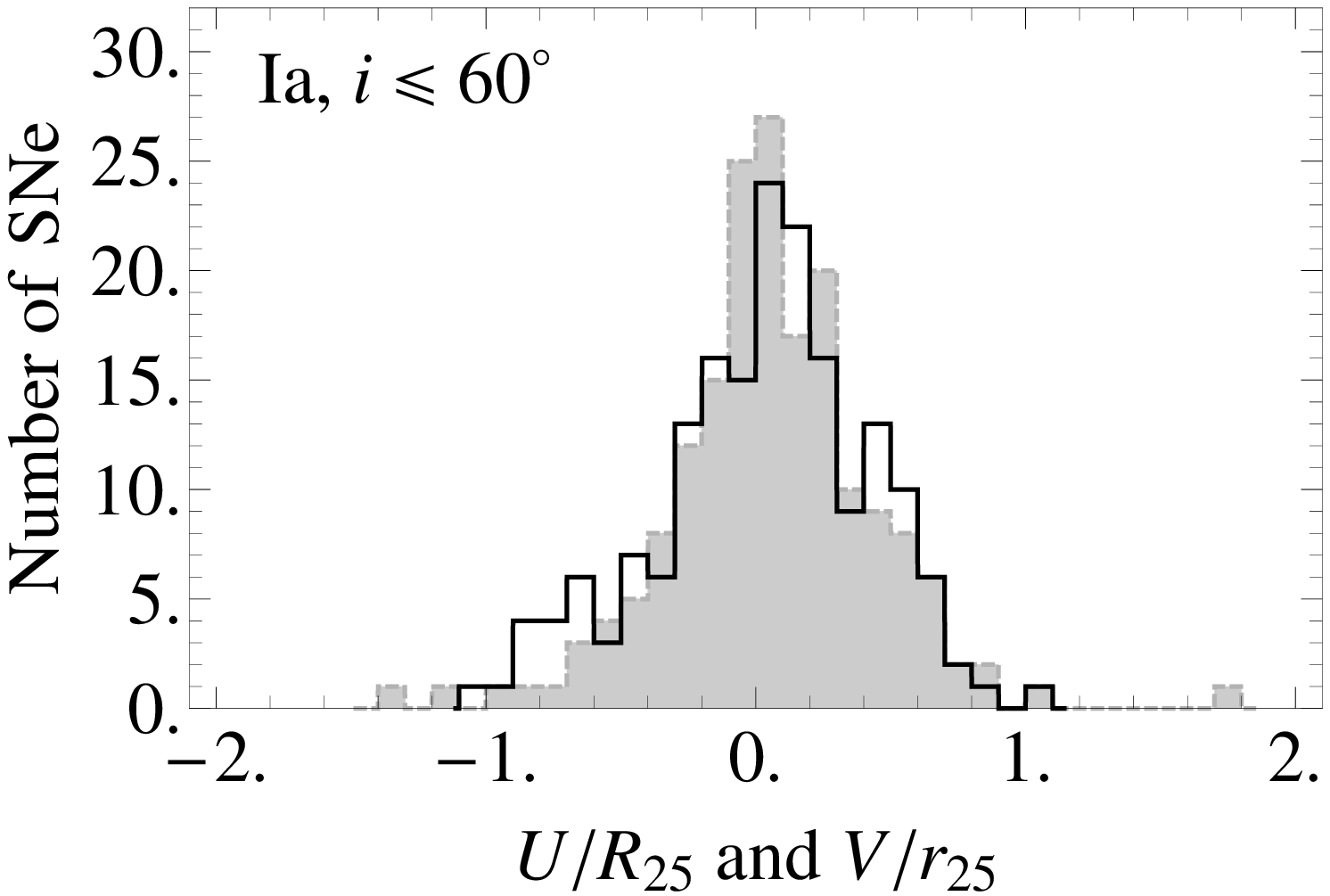} &
\includegraphics[width=0.22\hsize]{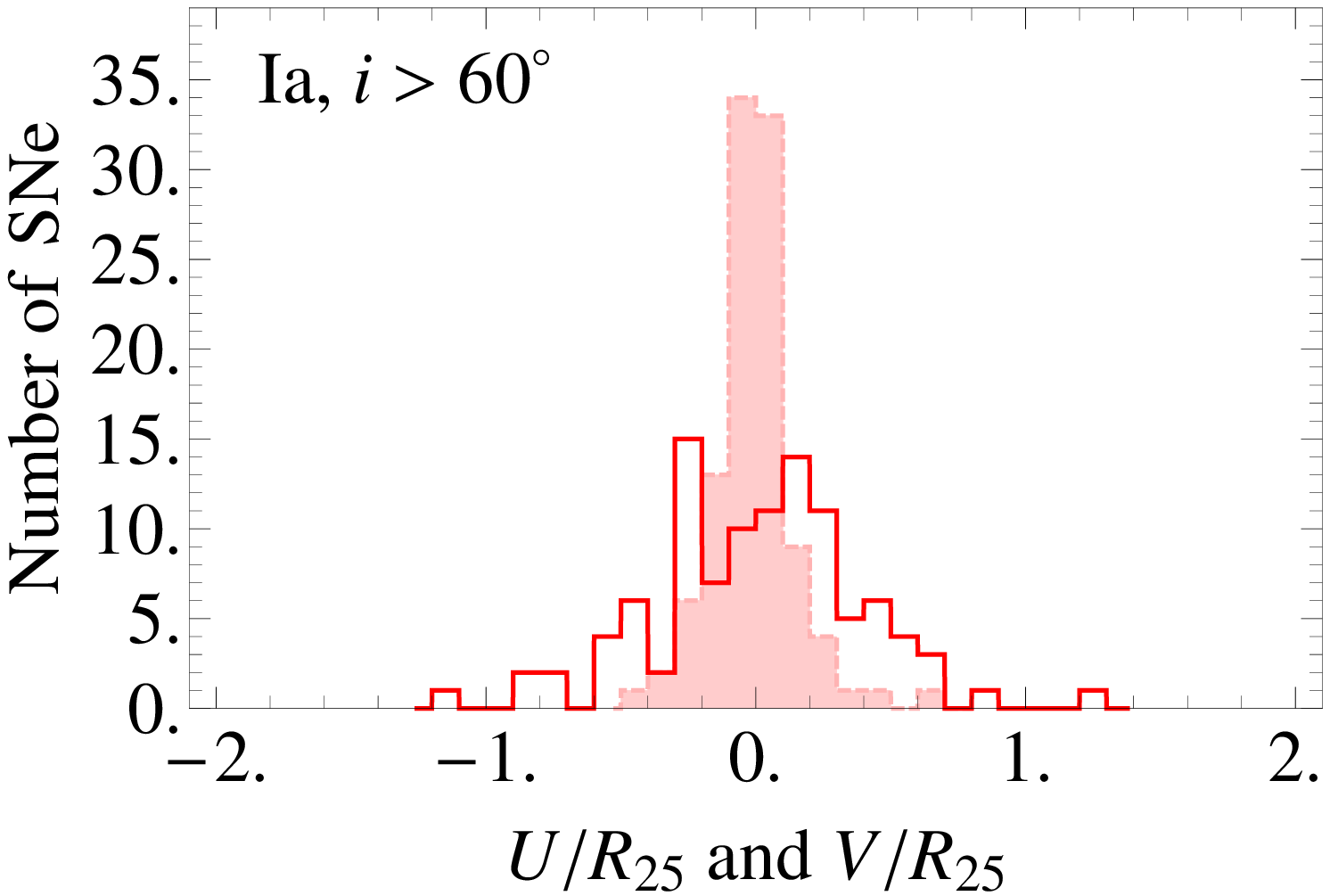} &
\includegraphics[width=0.22\hsize]{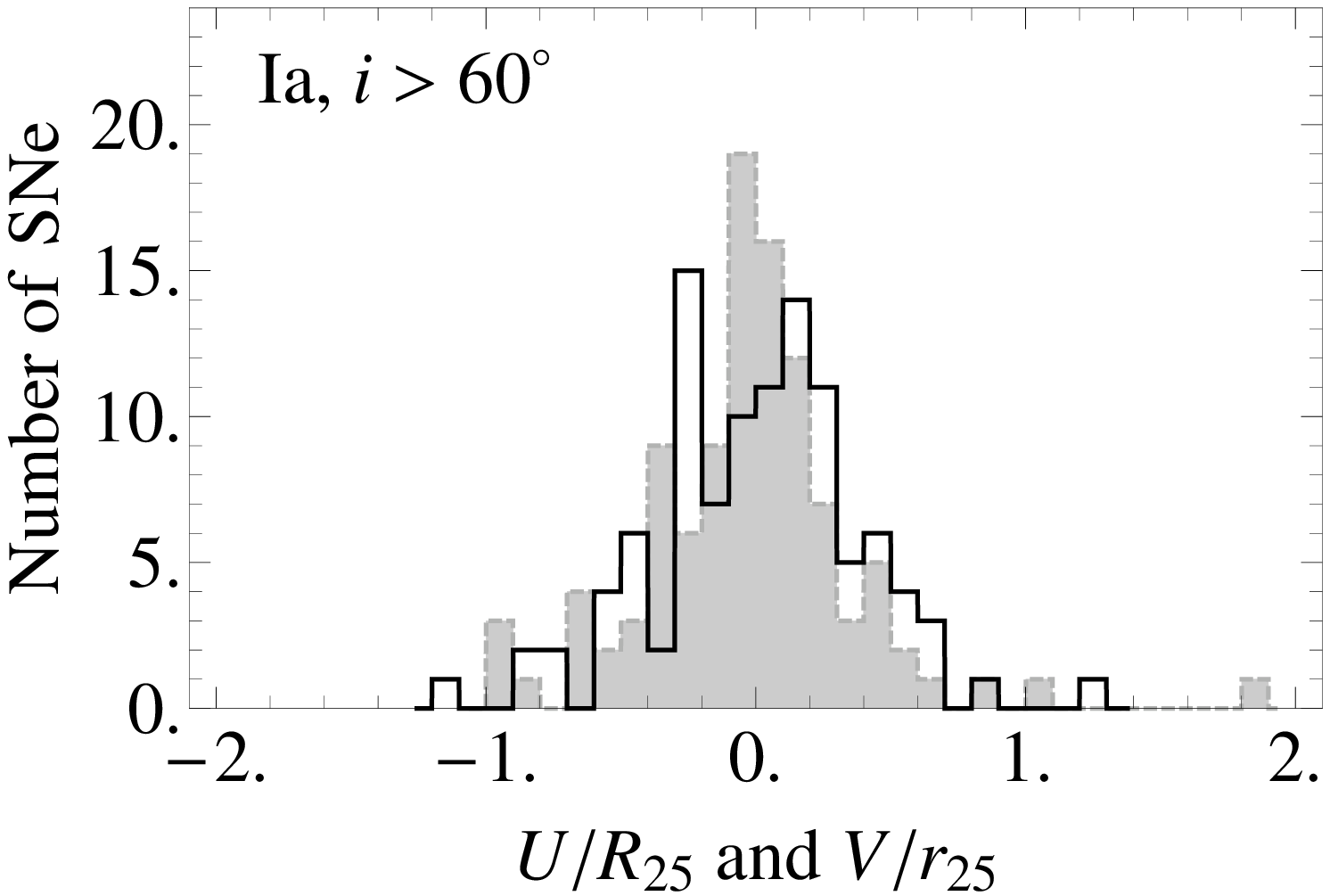}
\end{array}$
\end{center}
\begin{center}$
\begin{array}{@{\hspace{0mm}}l@{\hspace{0mm}}r@{\hspace{0mm}}}
\includegraphics[width=0.45\hsize]{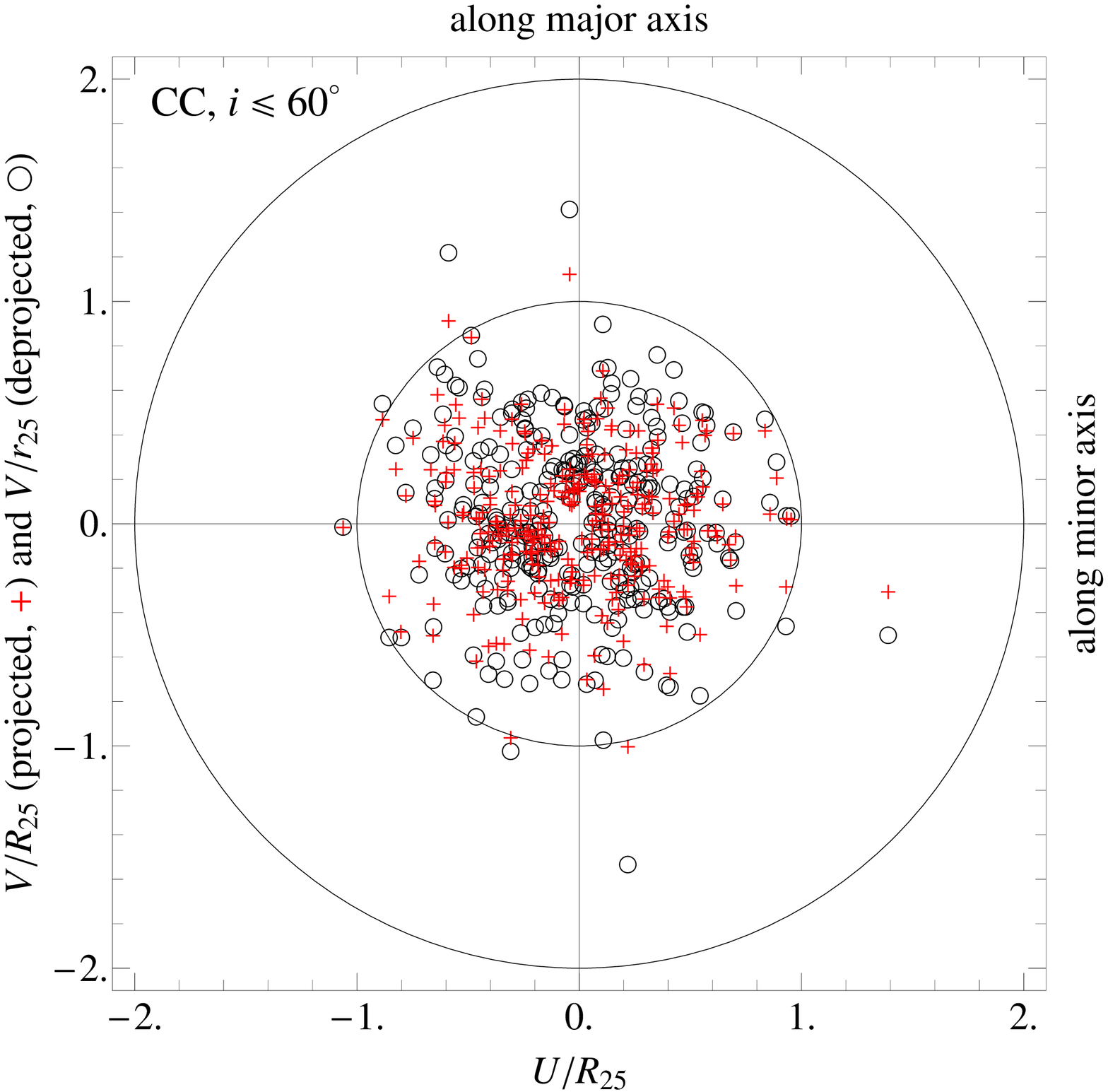} &
\includegraphics[width=0.45\hsize]{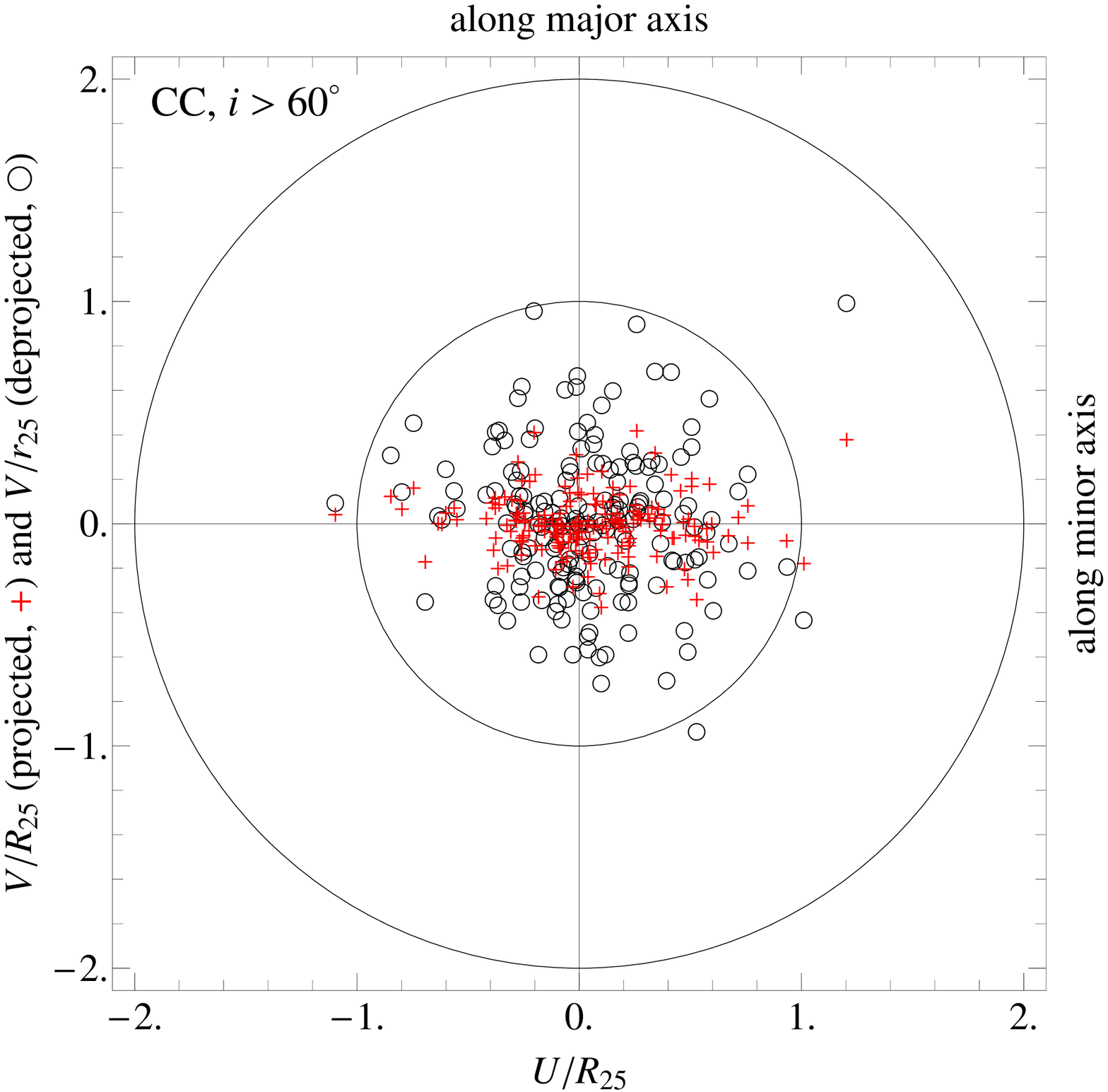}
\end{array}$
\end{center}
\begin{center}$
\begin{array}{@{\hspace{0mm}}r@{\hspace{0mm}}r@{\hspace{5mm}}r@{\hspace{0mm}}r@{\hspace{0mm}}}
\includegraphics[width=0.22\hsize]{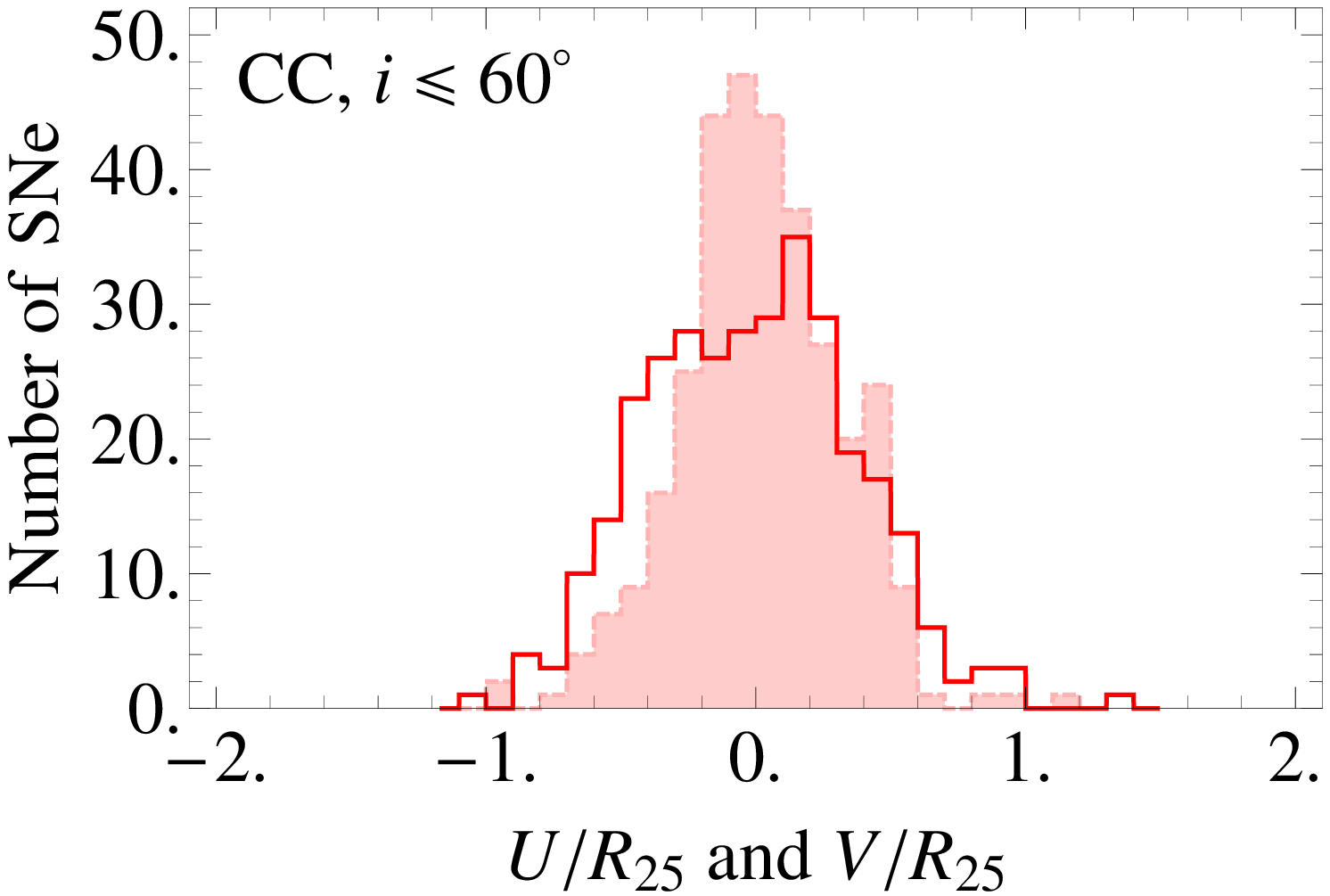} &
\includegraphics[width=0.22\hsize]{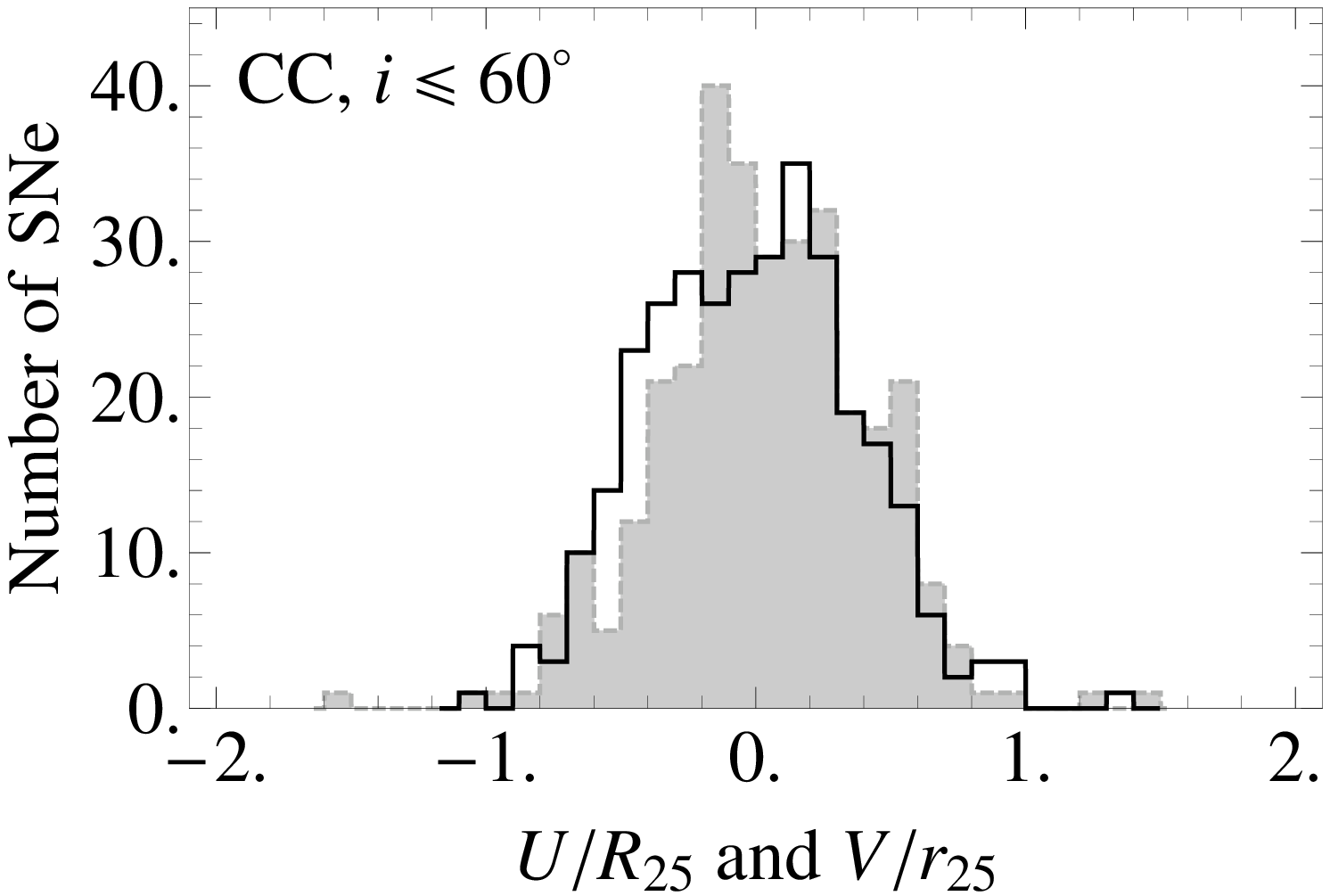} &
\includegraphics[width=0.22\hsize]{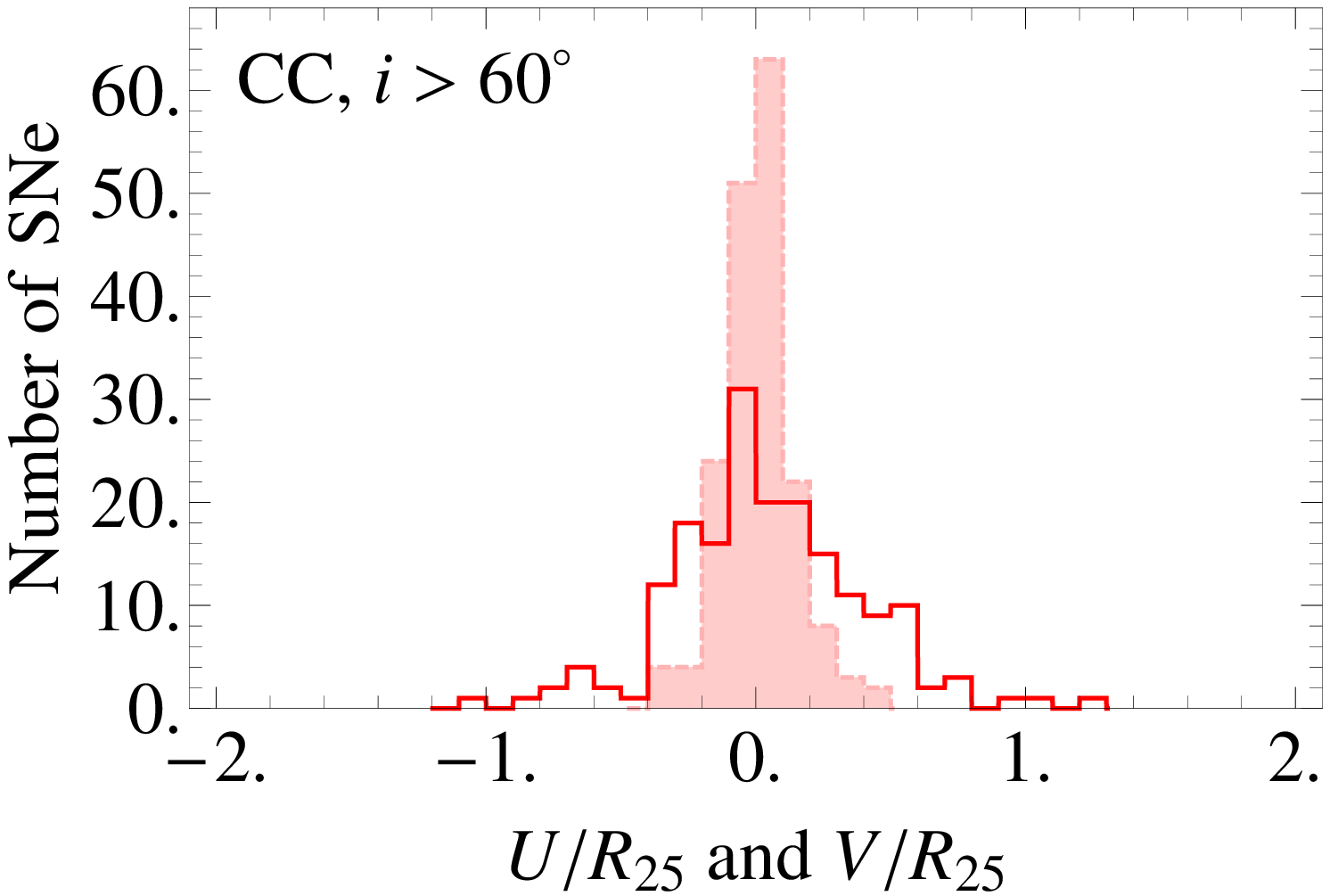} &
\includegraphics[width=0.22\hsize]{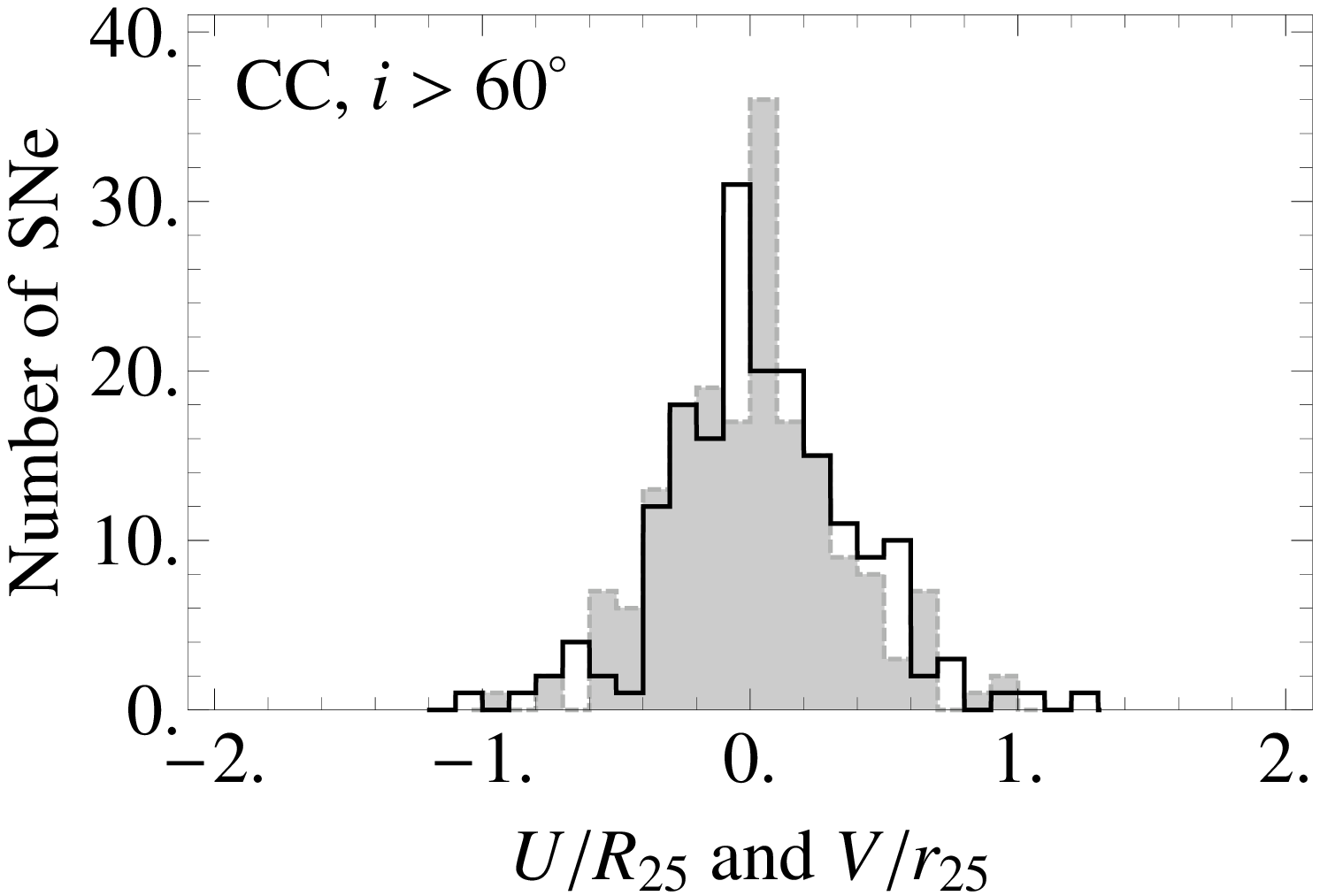}
\end{array}$
\end{center}
\caption{Distributions of normalized galactocentric distances of Type Ia and CC SNe along major ($U/R_{25}$)
         and minor ($V/R_{25}$ and $V/r_{25}$) axes in face-on ($i \leq 60^\circ$) and edge-on ($i > 60^\circ$)
         disc (S0--Sm) galaxies, where $r_{25}=R_{25}\,\cos i$ is the isophotal semiminor axis.
         The red crosses represent $U/R_{25}$ and $V/R_{25}$ pairs, while
         the black open circles show $U/R_{25}$ and $V/r_{25}$ pairs. The solid histograms are the distributions
         of $U/R_{25}$, while the shaded histograms are the distributions of $V/R_{25}$ and $V/r_{25}$.
         The two concentric circles are the host galaxy $R_{25}$ and $2R_{25}$ sizes.}
\label{hist_coord}
\end{figure*}

Fig.~\ref{hist_coord} shows (in red) the $V/R_{25}$ versus $U/R_{25}$ distributions in
face-on and edge-on S0--Sm galaxies, as well as their histograms.
The sample of SNe in face-on galaxies consists of
180 Ia and 320 CC SNe
(see Table~\ref{table_SN_morph}),
while the ancillary sample of SNe in edge-on hosts (not shown in Table~\ref{table_SN_morph})
includes 105 Ia and 181 CC SNe.
In disc galaxies, all types of SNe preferentially appear along the major $U$ axis
(see the standard deviations in Table~\ref{SNeindisc}).
This effect is stronger in edge-on galaxies.
In fact, the two-sample Kolmogorov--Smirnov (KS) and Anderson--Darling (AD)
tests,\footnote{{\footnotesize The two-sample AD test is more powerful than the KS test
\cite[][]{Engmann+11},
being more sensitive to differences in the tails of distributions.
Traditionally, we chose the threshold of 5 per cent for significance levels of the different tests.}}
shown in Table~\ref{SNeindisc},
indicate that the distributions of $U/R_{25}$ and $V/R_{25}$ are significantly
different, both for Type Ia and CC SNe
(except for Type Ia SNe in face-on hosts with the KS test).

\begin{table*}
  \centering
  \begin{minipage}{138mm}
  \caption{Comparison of the 2D spatial distributions of SNe among different subsamples of
           S0--Sm galaxies.}
  \tabcolsep 5pt
  \label{SNeindisc}
  \begin{tabular}{llccccccccccc}
  \hline
    \multicolumn{1}{c}{Subsamples}&\multicolumn{1}{c}{SN}&\multicolumn{5}{c}{$i \leq 60^\circ$}&&\multicolumn{5}{c}{$i > 60^\circ$}\\
  \cline{3-7} \cline{9-13}
    \multicolumn{1}{c}{(1) \, \, versus \, \, (2)}&&\multicolumn{1}{c}{$N_{\rm SN}$}&\multicolumn{1}{c}{$\sigma_{1}$}&\multicolumn{1}{c}{$\sigma_{2}$}&\multicolumn{1}{c}{$P_{\rm KS}$}&\multicolumn{1}{c}{$P_{\rm AD}$}&&\multicolumn{1}{c}{$N_{\rm SN}$}&\multicolumn{1}{c}{$\sigma_{1}$}&\multicolumn{1}{c}{$\sigma_{2}$}&\multicolumn{1}{c}{$P_{\rm KS}$}&\multicolumn{1}{c}{$P_{\rm AD}$}\\
  \hline
    $U/R_{25}$ versus $V/R_{25}$&Ia&180&0.394&0.299&0.217&\textbf{0.026}&&105&0.385&0.154&\textbf{0.001}&\textbf{0.000}\\
    $U/R_{25}$ versus $V/R_{25}$&CC&320&0.382&0.299&\textbf{0.003}&\textbf{0.002}&&181&0.349&0.132&\textbf{0.000}&\textbf{0.000}\\
    $U/R_{25}$ versus $V/r_{25}$&Ia&180&0.394&0.387&0.561&0.646&&105&0.385&0.390&0.174&0.321\\
    $U/R_{25}$ versus $V/r_{25}$&CC&320&0.382&0.384&0.120&0.162&&181&0.349&0.327&0.564&0.568\\
  \hline \\
  \end{tabular}
  \parbox{\hsize}{The $P_{\rm KS}$ and $P_{\rm AD}$ are the probabilities from two-sample KS and AD tests,
                  respectively, that the two distributions being compared (with respective standard
                  deviations $\sigma_1$ and $\sigma_2$) are drawn from the same
                  parent distribution.
                  The $P_{\rm KS}$ and $P_{\rm AD}$ are calculated using the calibrations by
                  \citet{Massey51} and \citet{Pettitt76}, respectively.
                  The statistically significant differences between the
                  distributions are highlighted in bold.}
  \end{minipage}
\end{table*}

It is worth mentioning that
Sa--Sm galaxies contain stellar populations of
different ages and host both Type Ia and CC SNe.
While S0 and S0/a galaxies,
which are mostly populated by old stars and have prominent bulges,
host Type Ia SNe. A tiny fraction of CC SNe has also been detected in galaxies classified as S0.
Nevertheless, in these rare cases, there is some evidence of residual star
formation in the SN hosts,
due to merging/accretion or interaction with close neighbours in the past
(e.g. \citealt[][]{2008A&A...488..523H}; Paper~\citetalias{2012A&A...544A..81H}).
The distribution of SNe types versus morphology of
the host galaxies, shown in Table~\ref{table_SN_morph}, is consistent with this picture.
In this sense, when selecting only Sa--Sm hosts, the
$P_{\rm KS}$ for Type Ia SNe in the face-on sample becomes significant (it is
reduced to 0.045), suggestive of a possible contribution from bulge SNe Ia
in S0--S0/a galaxies.
For Sa--Sm galaxies, the probabilities of KS and AD tests for the other subsamples
do not change the results of Table~\ref{SNeindisc}.

Performing simple calculations,
we find that the ratio of the mean values of $|V|$ and $|U|$ is 0.71 for Type Ia
and 0.75 for CC SNe in face-on, and, respectively, 0.35 and 0.36 in edge-on galaxies.
These numbers are very close to the mean values of the cosines of the inclinations
($\langle \cos \,\, i\rangle =0.74\pm0.01$ in face-on galaxies and
$0.31\pm0.02$ in edge-on galaxies -- where the uncertainties are the errors on the means),
thus supporting the fact that the vast majority of SNe in S0--Sm galaxies
are distributed within the stellar discs.
Similar results are found when only considering Sa--Sm galaxies.
This also suggests that the rate of SNe Ia in spiral galaxies is dominated by
a relatively young/intermediate progenitor population
\citep[e.g.][]{2005A&A...433..807M,2011Ap.....54..301H,2011MNRAS.412.1473L}.

For deprojected host discs,
in Fig.~\ref{hist_coord} we also show (in black)
the $V/r_{25}$ versus $U/R_{25}$ distributions
and their histograms for different samples,
where $r_{25}$ is the $g$-band 25th magnitude
isophotal semiminor axis ($r_{25}=R_{25} \, \cos \,\, i$).
In contrast to the previous normalization,
the KS and AD tests show that the distributions of $U/R_{25}$ and $V/r_{25}$ both for
Type Ia and CC SNe could be drawn from the same parent distribution (see Table~\ref{SNeindisc}).
In addition, the ratios of the mean values of
$|V|/r_{25}$ and $|U|/R_{25}$ in different samples are approximately equal to unity.
Thus, after correcting the host galaxies for inclination effects,
the distributions of SNe along major ($U/R_{25}$) and minor ($V/r_{25}$) axes turn to be equivalent
(see also the standard deviations in Table~\ref{SNeindisc}).
Hereafter, we restrict our analysis to the face-on sample
to minimize absorption and projection effects in the discs of
galaxies.\footnote{{\footnotesize The edge-on galaxies will be the sample for our
forthcoming paper with a different method of analysis.}}

We now adopt the oversimplified model where all SNe
are distributed on infinitely thin discs of the host galaxies.
While this is a reasonable assumption for SNe within spiral hosts, we shall extend this
assumption to S0--S0/a galaxies, for which many find a disc distribution of SNe
(e.g. \citealt{2008RMxAA..44..103O}; \citetalias{2009A&A...508.1259H} for CC and
\citealt{1999AJ....117.1185H,2000ApJ...542..588I} for Type Ia SNe).

In this thin-disc approximation, the corrected galactocentric radius
(in arcsec) of the SN in the host disc satisfies
\begin{equation}
R_{\rm SN}^2 = U^2+\left ({V\over \cos i} \right)^2 \ .\\
\label{RSN}
\end{equation}
We pay particular attention and discuss the cases when the contribution from the
bulge SNe Ia becomes apparent, especially in S0--S0/a hosts.

\subsection{Influence of bars and bulges on the radial distribution of SNe}
\label{resdiscus_sub2}

\begin{figure}
\begin{center}$
\begin{array}{@{\hspace{0mm}}c@{\hspace{0mm}}c@{\hspace{0mm}}c@{\hspace{0mm}}}
\includegraphics[width=0.95\hsize]{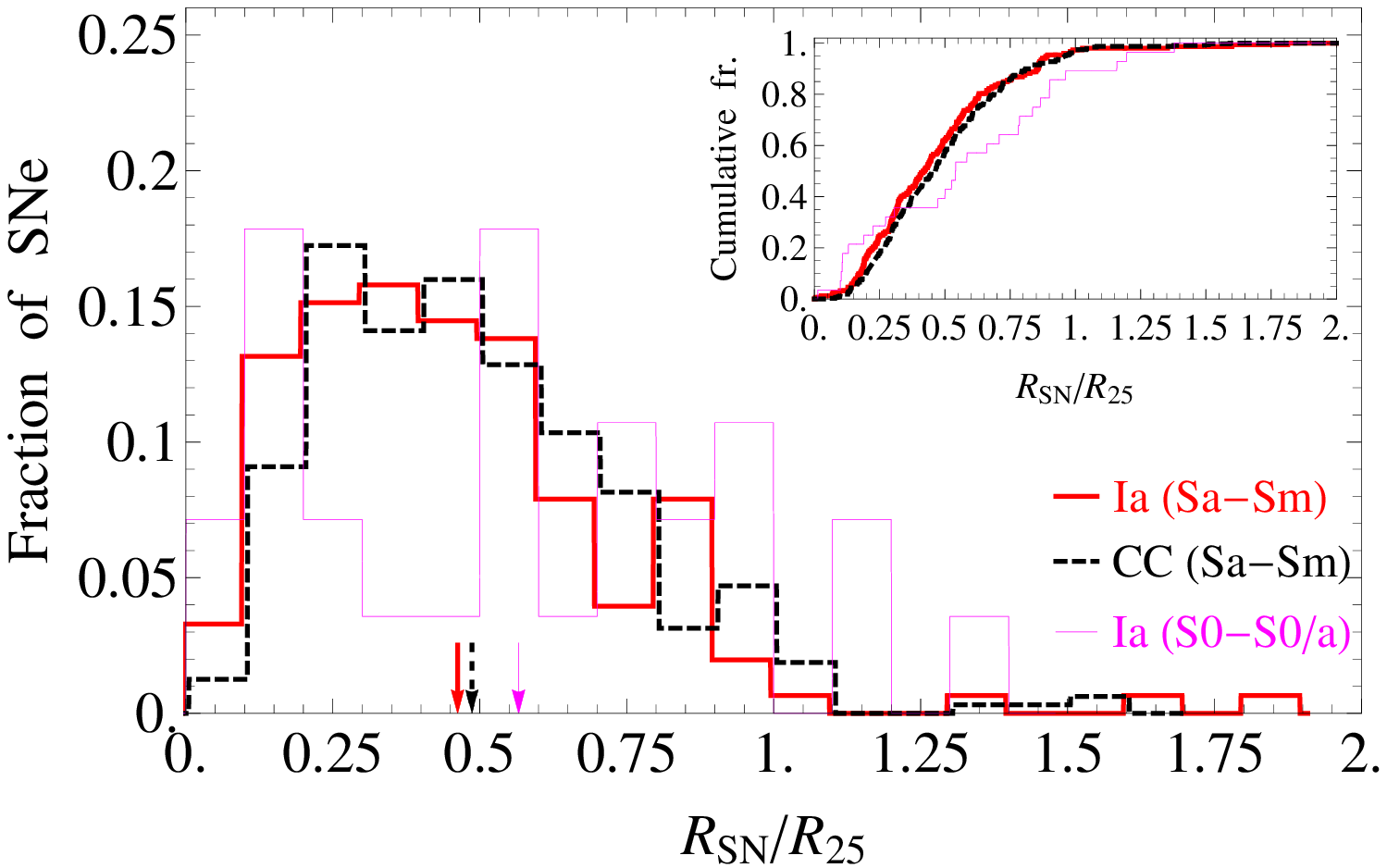}
\end{array}$
\end{center}
\begin{center}$
\begin{array}{@{\hspace{0mm}}c@{\hspace{0mm}}c@{\hspace{0mm}}c@{\hspace{0mm}}}
\includegraphics[width=0.95\hsize]{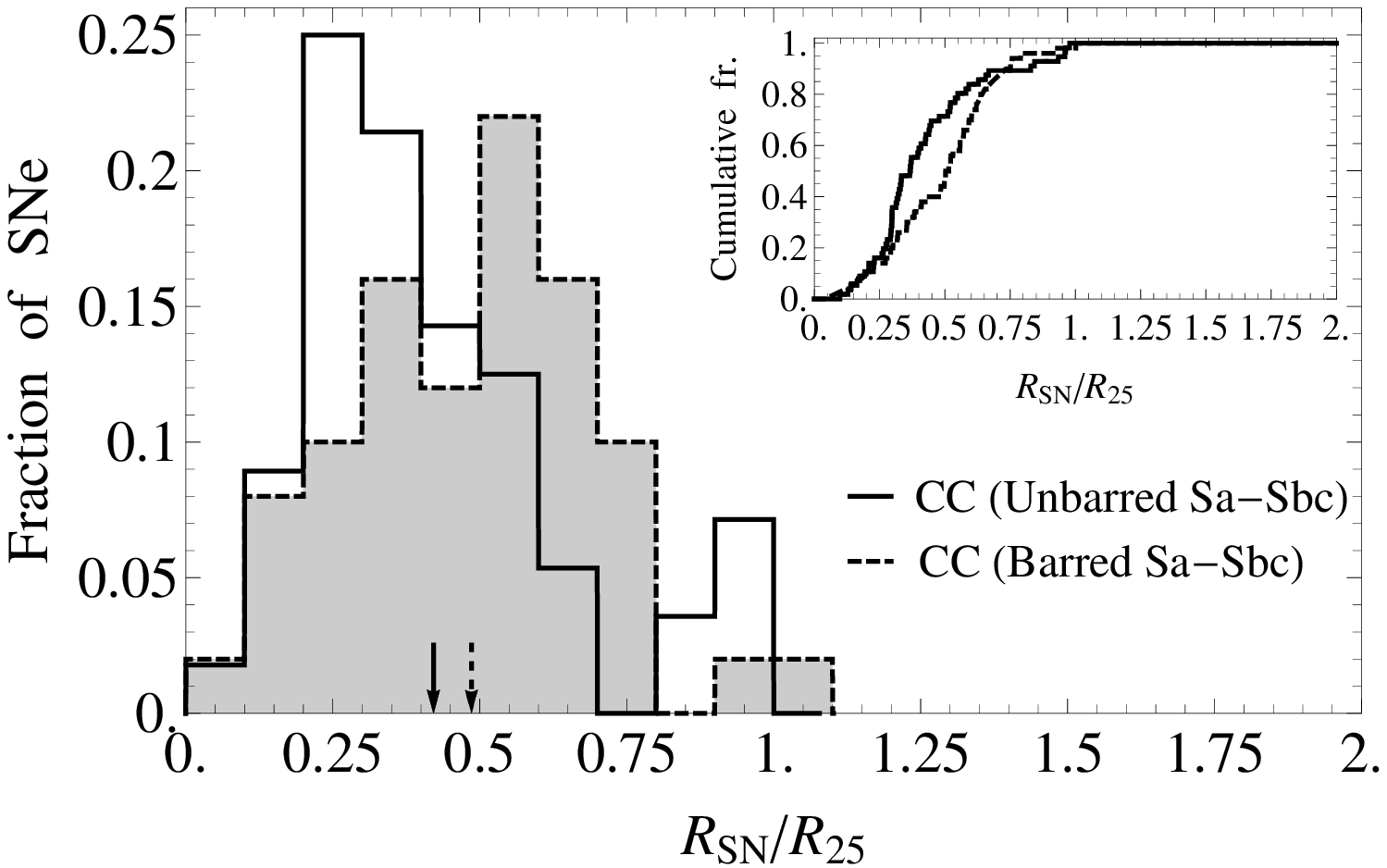}
\end{array}$
\end{center}
\begin{center}$
\begin{array}{@{\hspace{0mm}}c@{\hspace{0mm}}c@{\hspace{0mm}}c@{\hspace{0mm}}}
\includegraphics[width=0.95\hsize]{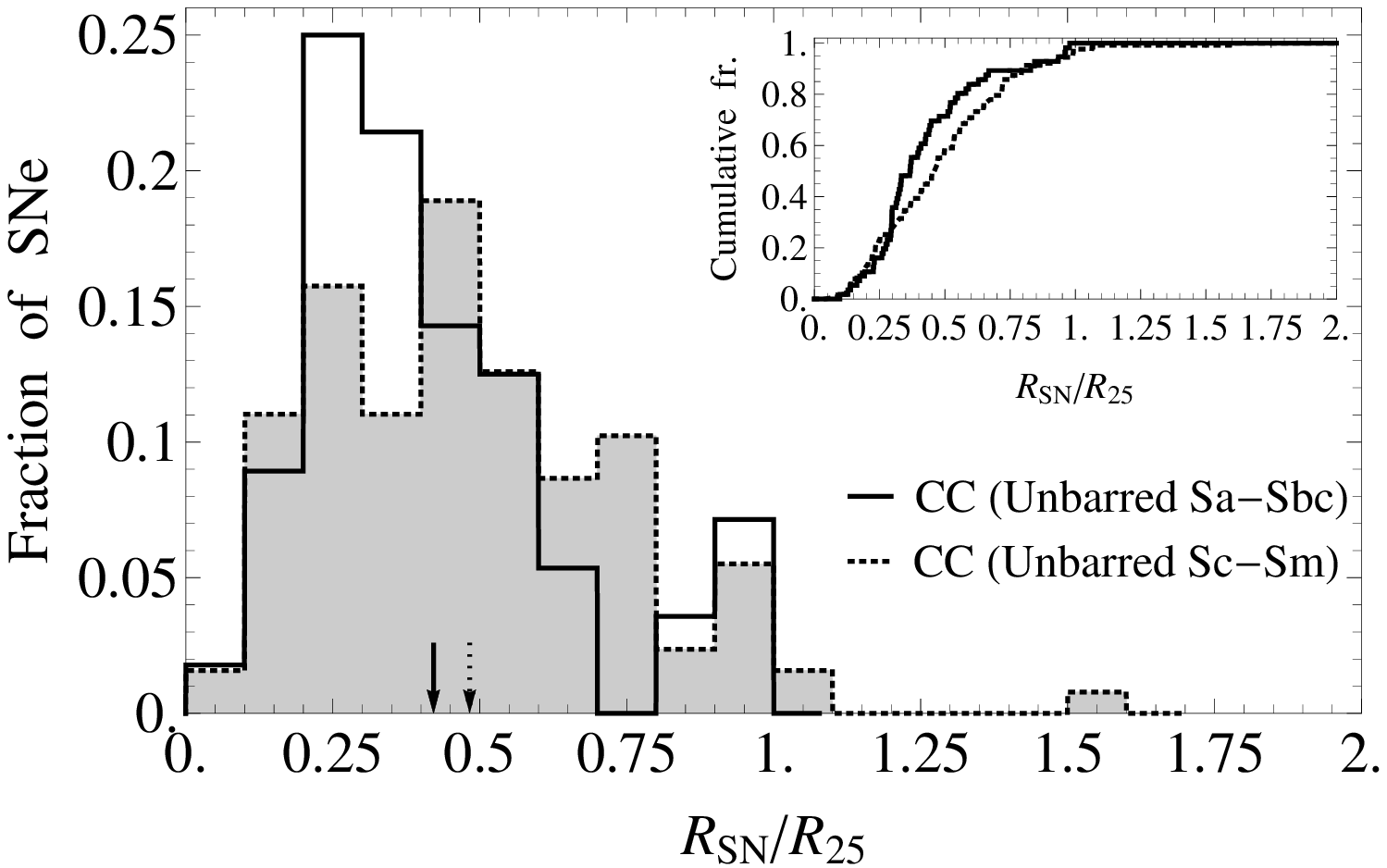}
\end{array}$
\end{center}
\caption{Upper panel: distributions of deprojected and normalized galactocentric radii
         ($\tilde{r}=R_{\rm SN}/R_{25}$) of Type Ia SNe in S0--S0/a and Sa--Sm hosts,
         as well as CC SNe in Sa--Sm galaxies.
         Middle panel: the distributions of CC SNe in barred and unbarred Sa--Sbc hosts.
         Bottom panel: the distributions of CC SNe in unbarred Sa--Sbc and Sc--Sm galaxies.
         The insets present the cumulative distributions of SNe.
         The mean values of the distributions are shown by arrows.}
\label{hist_distr}
\end{figure}

The upper panel of Fig.~\ref{hist_distr} compares the distributions of deprojected,
normalized galactocentric radii ($\tilde{r}=R_{\rm SN}/R_{25}$) of
Type Ia and CC SNe in S0--S0/a and Sa--Sm galaxies.
In this panel, we see an initial rise of SNe number
as a function of $\tilde{r}$ and a negative radial gradient outside
the maximum, suggestive of the exponential surface brightness distribution of
stellar discs, which we will study in more detail in Section~\ref{resdiscus_sub3}.
In previous studies, based on smaller samples,
similar deprojected distributions of SNe were already obtained by
normalizing them to the radii of hosts at some fixed surface brightness isophot
(\citealt{1963PASP...75..123J,1975A&A....44..267B,
1975PASJ...27..411I,1977MNRAS.178..693V,1986Ap&SS.122..343B,1989A&A...217...79T,
1990A&A...239...63P,1995A&A...297...49P,1995A&A...301..666L,2000ApJ...542..588I,
2001MNRAS.328.1181N,2002MNRAS.331L..25B,2005AJ....129.1369P,2008Ap.....51...69H,
2009MNRAS.399..559A}; \citetalias{2009A&A...508.1259H};
\citealt{2010MNRAS.405.2529W,2013Ap&SS.347..365N,2013Sci...340..170W}).

The observed numbers of SNe at $\tilde{r} \lesssim 0.2$ indicate
that different SN searches fail to discover objects at or near the centre of
the surveyed galaxies \cite[e.g.][]{1997A&A...322..431C,1999AJ....117.1185H}.
In an earlier study, \cite{1979A&A....76..188S} noted that this effect is stronger for
distant host galaxies relative to nearer ones.
Then it became apparent that this \emph{Shaw effect} is important for deep
photographic searches and negligible for visual and CCD searches
\cite[e.g.][]{1997A&A...322..431C,2000ApJ...530..166H}.
Presently, SN searches are conducted only with CCD cameras and SNe are
discovered via image subtraction, so the discrimination against SNe
occurring near the bright nuclei of galaxies is less strong.
Nevertheless, an area with a radius of a few pixels centred on
every galaxy nucleus is usually excluded during a search,
because galactic nuclei often suffer imperfect image subtraction and
introduce many false sources \cite[e.g.][]{2011MNRAS.412.1419L}.
Another difficulty is that extinction by dust in host galaxy discs,
depending on inclination, can affect the radial distributions of SNe,
particularly in the nuclear region
\cite[e.g.][]{1997ApJ...483L..29W,1998ApJ...502..177H,2015MNRAS.446.3768H}.

To check the possible influences of these selection effects on
the radial distribution of SNe, we compare, in Table~\ref{SNeIaCC_KS_AD},
the radial distributions of SNe in different pairs of subsamples where these
selection effects may produce different radial SN distributions.
For this, we performed two-sample KS and AD tests between the pairs of radial
SN distributions.
These tests show that the radial
distributions of Type Ia and CC SNe in our sample are not affected by
distance or inclination of the host galaxy, nor by
the SN discovery epoch (photographic/CCD searches).

We now compare, in Table~\ref{diffSNe_KS_AD}, the distributions of the
normalized deprojected radii for pairs of subsamples that should not
be affected by selection effects.
We see no statistically significant differences between the radial
distributions of Type Ia and CC SNe in all
the subsamples of Sa--Sm galaxies.
In contrast, the cumulative radial distributions of SNe Ia in S0--S0/a and Sa--Sm galaxies
apparently deviate from one another (as seen in the AD statistic but only
very marginally so in the KS statistic). Fig.~\ref{hist_distr}, indeed shows
signs of a bimodal radial distribution for SNe~Ia in S0--S0/a galaxies.

\begin{table*}
  \centering
  \begin{minipage}{167mm}
  \caption{Comparison of the normalized, deprojected radial distributions of
           SNe among different pairs of subsamples possibly
           affected by selection effects.}
  \tabcolsep 3pt
  \label{SNeIaCC_KS_AD}
  \begin{tabular}{lclrrcccrrcc}
  \hline
    &&&\multicolumn{4}{c}{Ia}&&\multicolumn{4}{c}{CC}\\
  \cline{4-7} \cline{9-12}
    \multicolumn{1}{l}{Subsample~1}&&\multicolumn{1}{l}{Subsample~2}&\multicolumn{1}{c}{$N_{\rm SN}$(1)}&\multicolumn{1}{c}{$N_{\rm SN}$(2)}&\multicolumn{1}{c}{$P_{\rm KS}$}&\multicolumn{1}{c}{$P_{\rm AD}$}&&\multicolumn{1}{c}{$N_{\rm SN}$(1)}&\multicolumn{1}{c}{$N_{\rm SN}$(2)}&\multicolumn{1}{c}{$P_{\rm KS}$}&\multicolumn{1}{c}{$P_{\rm AD}$}\\
  \hline
    $0 < D {\rm (Mpc)} \leq 60$ (S0--S0/a) &versus&$60 < D {\rm (Mpc)} \leq 100$ (S0--S0/a) &10&18&0.566&0.416&&0&1&--&--\\
    $0 < D {\rm (Mpc)} \leq 60$ (Sa--Sm) &versus&$60 < D {\rm (Mpc)} \leq 100$ (Sa--Sm) &69&83&0.692&0.620&&168&151&0.683&0.465\\
    $0^\circ \leq i \leq 40^\circ$ (S0--S0/a) &versus&$40^\circ < i \leq 60^\circ$ (S0--S0/a) &11&17&0.973&0.883&&0&1&--&--\\
    $0^\circ \leq i \leq 40^\circ$ (Sa--Sm) &versus&$40^\circ < i \leq 60^\circ$ (Sa--Sm) &64&88&0.190&0.412&&134&185&0.938&0.821\\
    $t \leq 2000 \, {\rm yr}$ (S0--S0/a) &versus&$t > 2000 \, {\rm yr}$ (S0--S0/a) &6&22&0.348&0.560&&0&1&--&--\\
    $t \leq 2000 \, {\rm yr}$ (Sa--Sm) &versus&$t > 2000 \, {\rm yr}$ (Sa--Sm) &58&94&0.997&0.857&&108&211&0.517&0.635\\
  \hline \\
  \end{tabular}
  \parbox{\hsize}{Each pair of the subsamples is selected, if possible, to include comparable numbers of
                  SNe in both subsamples.}
  \end{minipage}
\end{table*}
\begin{table*}
  \centering
  \begin{minipage}{113mm}
  \caption{Comparison of the normalized, deprojected radial distributions of SNe
           among different pairs of subsamples.}
  \tabcolsep 5pt
  \label{diffSNe_KS_AD}
  \begin{tabular}{llrcllrcc}
  \hline
    \multicolumn{3}{c}{Subsample~1}&&\multicolumn{3}{c}{Subsample~2}&&\\
  \cline{1-3} \cline{5-7}
    \multicolumn{1}{l}{Host}&\multicolumn{1}{c}{SN}&\multicolumn{1}{c}{$N_{\rm SN}$}&&\multicolumn{1}{l}{Host}&\multicolumn{1}{c}{SN}&\multicolumn{1}{c}{$N_{\rm SN}$}&\multicolumn{1}{c}{$P_{\rm KS}$}&\multicolumn{1}{c}{$P_{\rm AD}$}\\
  \hline
    S0--S0/a&Ia&28&versus&Sa--Sm&Ia&152&0.099&\textbf{0.020}\\
  \hline
    Sa--Sm&Ia&152&versus&Sa--Sm&CC&319&0.568&0.259\\
    Sa--Sm (barred)&Ia&73&versus&Sa--Sm (barred)&CC&136&0.483&0.378\\
    Sa--Sm (unbarred)&Ia&79&versus&Sa--Sm (unbarred)&CC&183&0.349&0.342\\
    Sa--Sm (barred)&Ia&73&versus&Sa--Sm (unbarred)&Ia&79&0.192&0.313\\
    Sa--Sm (barred)&CC&136&versus&Sa--Sm (unbarred)&CC&183&0.060&0.076\\
  \hline
    Sa--Sbc&Ia&79&versus&Sa--Sbc&CC&106&0.683&0.751\\
    Sa--Sbc (barred)&Ia&39&versus&Sa--Sbc (barred)&CC&50&0.528&0.627\\
    Sa--Sbc (unbarred)&Ia&40&versus&Sa--Sbc (unbarred)&CC&56&0.702&0.473\\
    Sa--Sbc (barred)&Ia&39&versus&Sa--Sbc (unbarred)&Ia&40&0.242&0.617\\
    Sa--Sbc (barred)&CC&50&versus&Sa--Sbc (unbarred)&CC&56&\textbf{0.008}&\textbf{0.028}\\
  \hline
    Sc--Sm&Ia&73&versus&Sc--Sm&CC&213&0.424&0.276\\
    Sc--Sm (barred)&Ia&34&versus&Sc--Sm (barred)&CC&86&0.670&0.575\\
    Sc--Sm (unbarred)&Ia&39&versus&Sc--Sm (unbarred)&CC&127&0.340&0.355\\
    Sc--Sm (barred)&Ia&34&versus&Sc--Sm (unbarred)&Ia&39&0.503&0.615\\
    Sc--Sm (barred)&CC&86&versus&Sc--Sm (unbarred)&CC&127&0.581&0.276\\
  \hline
    Sa--Sbc&Ia&79&versus&Sc--Sm&Ia&73&0.850&0.956\\
    Sa--Sbc&CC&106&versus&Sc--Sm&CC&213&0.183&0.156\\
    Sa--Sbc (barred)&Ia&39&versus&Sc--Sm (barred)&Ia&34&0.992&0.967\\
    Sa--Sbc (barred)&CC&50&versus&Sc--Sm (barred)&CC&86&0.489&0.278\\
    Sa--Sbc (unbarred)&Ia&40&versus&Sc--Sm (unbarred)&Ia&39&0.822&0.961\\
    Sa--Sbc (unbarred)&CC&56&versus&Sc--Sm (unbarred)&CC&127&\textbf{0.033}&0.067\\
  \hline \\
  \end{tabular}
  \parbox{\hsize}{The statistically significant differences between the
                  distributions are highlighted in bold.}
  \end{minipage}
\end{table*}

Naturally, one expects that among disc galaxies, bulge SNe Ia
should have highest contribution to the whole distribution of SNe Ia
in S0--S0/a subsample, because among disc galaxies the ratio of bulge luminosity over
disc luminosity (or B/D) is highest in the S0--S0/a subsample
\citep[e.g.][]{1975A&A....44..363Y,1986ApJ...302..564S,2009ApJ...705..245O,2011A&A...532A..75D}.
In the case of Sa--Sm galaxies, both Type Ia and CC SNe dominate in the Sbc--Sc morphological bin
(see Table~\ref{table_SN_morph}), where the B/D ratio is significantly lower than that
in S0--S0/a galaxies \cite[e.g.][]{2009ApJ...705..245O,2011A&A...532A..75D}.
Moreover, the radial distributions of Type Ia and CC SNe in all the
Sa--Sm subsamples can be drawn from the same parent distribution
(see the $P$-values in Table~\ref{diffSNe_KS_AD}) supporting the fact that in these galaxies
SNe Ia mostly exploded in the disc component where all CC SNe also occur.
Thus, the apparent deviation of the radial distribution of Type Ia SNe in S0--S0/a galaxies
from that in Sa--Sm hosts (upper panel of Fig.~\ref{hist_distr})
is attributed to the contribution by SNe Ia from
the bulge component of S0--S0/a galaxies.

Another interesting results stands out in Table~\ref{diffSNe_KS_AD}.
The radial distributions of CC SNe in barred versus unbarred Sa--Sbc galaxies
are inconsistent (middle panel of Fig.~\ref{hist_distr}),
while for Type Ia SNe the radial distributions between barred and unbarred
Sa--Sbc galaxies are not significantly different.
At the same time, the radial distributions of both Type Ia and CC SNe in Sc--Sm galaxies
are not affected by bars.

Interestingly, \citet{2009A&A...501..207J} discovered that
several early-type barred spiral galaxies that had strong dips in their radial
H$\alpha$ line emission profiles, near
$0.25\,R_{25}$.\footnote{\footnotesize{The dip in H$\alpha$ occurs at
one-quarter of the isophotal radius corresponding to a surface magnitude of
$\mu_R=24\,\rm mag\,arcsec^{-2}$, which in turn is roughly equal to $R_{25}$.}}
This region is roughly where their unbarred counterparts host the strongest star formation.
The H$\alpha$ emission from barred Sb--Sbc galaxies showed central components and
concentrations of star formation at or just outside the bar-end radius.
Except for the central component, where we have difficulties to discover SNe,
the picture of \citeauthor{2009A&A...501..207J} is very similar to that we observe for CC SNe in
barred and unbarred Sa--Sbc galaxies
(see the middle panel of Fig.~\ref{hist_distr} and fig.~8 in \citeauthor{2009A&A...501..207J}).
The region between the central component and bar-end radius of H$\alpha$ emission profiles
is termed the `star formation desert' (SFD) by \citet{2015MNRAS.450.3503J}.
In addition, they noted that the SFD had significant continuum emission in
the $R$-band and even showed line emission not consistent with expectations
from star formation, but most probably from an old stellar population.

Taking into consideration that the distributions of Type Ia and CC SNe,
respectively, trace the distributions of $R$-band continuum emission/stellar mass
and H$\alpha$ emission/star formation \citep[e.g.][]{2006A&A...453...57J,2009MNRAS.399..559A},
we compare the inner ($\tilde{r} \leqslant 0.3$) fractions of different SNe ($F_{\rm SN}$) in
barred and unbarred host galaxies (see Table~\ref{table0.3ratios}).
This inner region is selected because in Sa--Sbc subsample the SNe dominate in
Sb--Sbc galaxies (see Table~\ref{table_SN_morph}), where the mean SFD region has outer
radius $\sim0.3$ of the optical radius
(see fig.~5 in \citealt{2009A&A...501..207J}).\footnote{{\footnotesize Ideally,
instead of using the $\tilde{r} \leqslant 0.3$ region,
we could perform bar length measurements for each of barred galaxy in the sample
to obtain the fractions of different SNe in the radial range swept by bars.
However, this is beyond the scope of this paper,
and will be the subject of a future paper.}}

Table~\ref{table0.3ratios} shows that the inner $\tilde{r} \leqslant 0.3$ fractions of
Type Ia SNe in Sa--Sbc and Sc--Sm hosts are not statistically different
between barred and unbarred galaxies.
The same situation holds true for CC SNe in Sc--Sm galaxies.
However, the inner fraction of CC SNe is significantly lower in
barred Sa--Sbc galaxies compared with their unbarred counterparts.
It is important to note that the inner fractions of Type Ia and CC SNe
are not statistically different one from another when the same morphological
and barred/unbarred categories are selected.

\begin{table}
  \centering
  \begin{minipage}{83mm}
  \caption{Fractions of inner ($R_{\rm SN} \leq 0.3\,R_{25}$) SNe in
           barred and unbarred host galaxies.}
  \tabcolsep 5pt
  \label{table0.3ratios}
  \begin{tabular}{llrccrcc}
  \hline
    &&\multicolumn{2}{c}{Barred}&&\multicolumn{2}{c}{Unbarred}&\\
  \cline{3-4} \cline{6-7}
    \multicolumn{1}{l}{Host}&\multicolumn{1}{l}{SN}&\multicolumn{1}{c}{$N_{\rm SN}$}&\multicolumn{1}{c}{$F_{\rm SN}$}&&\multicolumn{1}{c}{$N_{\rm SN}$}&\multicolumn{1}{c}{$F_{\rm SN}$}&\multicolumn{1}{c}{$P_{\rm B}$}\\
  \hline
    Sa--Sbc&Ia&39&$0.26_{-0.04}^{+0.05}$&&40&$0.38_{-0.05}^{+0.05}$&0.170\\
    Sa--Sbc&CC&50&$0.20_{-0.03}^{+0.04}$&&56&$0.36_{-0.04}^{+0.04}$&\textbf{0.044}\\
    Sc--Sm&Ia&34&$0.26_{-0.05}^{+0.05}$&&39&$0.36_{-0.05}^{+0.05}$&0.240\\
    Sc--Sm&CC&86&$0.26_{-0.03}^{+0.03}$&&127&$0.28_{-0.02}^{+0.02}$&0.741\\
  \hline \\
  \end{tabular}
  \parbox{\hsize}{The standard deviations of the fractions are calculated using the approach of
                  \citet{2011PASA...28..128C}.
                  The significance value $P_{\rm B}$ is calculated using Barnard's exact test \citep{1945Natur.156..177B},
                  which compares the pairs of numbers rather than the fractions.
                  The statistically significant difference between the fractions
                  is highlighted in bold.}
  \end{minipage}
\end{table}

The results of Table~\ref{table0.3ratios} agree quite well with the findings of
\citet{2015MNRAS.450.3503J}
that barred galaxies of earlier Hubble types have substantially suppressed star formation
\citep[see also][]{2009A&A...501..207J}, hence the lack of CC SNe in the
inner radial range swept by the strong bars of Sa--Sbc galaxies.
On the other hand, this region is not `forbidden' to Type Ia SNe,
because these SNe originate from an older stellar population also located in the bulge.
CC SNe in barred Sc--Sm galaxies can appear in the inner regions
because the effect of star formation suppressing by bars is not seen in late-type barred galaxies
\citep[][]{2009A&A...501..207J,2015MNRAS.450.3503J}.
Thus, we see that bars of host galaxies affect the radial distributions of SNe,
at least the CC ones, in the stellar discs of early-type galaxies.

According to Table~\ref{diffSNe_KS_AD} and
the bottom panel of Fig.~\ref{hist_distr}, we also see that
the radial distribution of CC SNe in unbarred Sa--Sbc galaxies is more centrally peaked
and inconsistent with that in unbarred Sc--Sm hosts (as seen in the KS statistic but only
marginally so in the AD statistic).
In contrast, the radial distribution of Type Ia SNe in unbarred galaxies
is unaffected by host morphology.

The different radial distributions of SNe in unbarred spirals
can be explained by the strong dependence of massive star formation distribution
in the discs on the morphological type of galaxies.
In particular, for S0/a--Im galaxies \citet[][]{2009A&A...501..207J}
studied the distribution of H$\alpha$ and $R$-band concentration
indices (C30)\footnote{\footnotesize{The C30 index
is the ratio of the flux within 0.3 times the optical radius and the total flux,
which provides a simple measure of the observed radial distribution of
the luminosity of a galaxy in a specified bandpass.}}
as a function of Hubble type.
They found a strong correlation of H$\alpha$ C30 index with Hubble sequence,
with the early-type having about two times higher H$\alpha$ C30 indices in
comparison with the late-type galaxies.
The same is true for $R$-band C30 indices, but with less difference between
early- and late-type galaxies \citep[see fig~1. in][]{2009A&A...501..207J}.
Moreover, the authors showed that the unbarred galaxies have more centrally
concentrated H$\alpha$ emission than do their strongly barred counterparts.
Note that, in some cases, the C30 indices may be biased by
Active Galactic Nuclei (AGN) emission from the very central region
($<0.2 R_{25}$) of galaxies.
However, fig.~8 in \citet[][]{2009A&A...501..207J} shows that
for unbarred early-type galaxies the trends above are virtually unaffected
when the inner regions ($<0.2 R_{25}$) are not considered.
Since Type Ia SNe are less tightly connected to the H$\alpha$ emission
of the explosion site \citep[e.g.][]{2012A&A...545A..58S,2014A&A...572A..38G},
the radial distribution of SNe Ia in Sa--Sm hosts
is not strongly affected by the morphology.

It is important to note that, when selecting Sa--Sbc and Sc--Sm morphologies
without splitting between barred and unbarred subsamples, all the significant
differences in the radial distributions of SNe are washed out (see Table~\ref{diffSNe_KS_AD}).
Therefore, the lack of significant differences in the radial distributions of
CC SNe as a function of the morphological type of host galaxies
presented in the earlier studies
(e.g. \citealt{1992A&A...264..428B,1995A&A...301..666L};
\citealt*{2004AstL...30..729T};
\citealt{2005AJ....129.1369P,2008Ap.....51...69H};
\citetalias{2009A&A...508.1259H};
\citealt{2010MNRAS.405.2529W}) is a consequence of these samples
mixing barred and unbarred galaxies with different levels of mixing of the
stellar populations.

To exclude the effects of (1) bars
(two distinct types of bars: strong bars, which are more common in early-type discs,
and weak bars, which are frequently found in late-type spirals;
see fig.~5 of Paper~\citetalias{2014MNRAS.444.2428H}) and (2)
the morphological differences between Type Ia and CC SNe hosts
(the mean morphological type of SNe Ia host galaxies is earlier than that of
the CC SNe hosts, as can be deduced from the numbers in our
Table~\ref{table_SN_morph}; see also
fig.~2 and table~11 of Paper~\citetalias{2014MNRAS.444.2428H}),
we repeat the analysis of the inner ($\tilde{r} \leqslant 0.3$) fractions of
SNe, restricting to unbarred morphological bins (Sbc and Sb--Sbc).
We carry out this analysis because the bulge stars of Sb--Sbc galaxies are
typically located in our chosen inner
region,\footnote{\footnotesize{Fig.~23 of \citet{2014MNRAS.443..874B}
indicates that the bulge half-light radius, $R_{\rm bulge}$, is roughly one-quarter of the
disc scale length, which for typical central disc surface brightness
amounts to $R_{\rm bulge} < R_{25}/10$ (using the relation between disc scale
length and $R_{25}$ given in equation 3 of \citetalias{2009A&A...508.1259H}).}}
which enables a better estimation of
the possible contribution of bulge SNe Ia on top of that of the inner disc population.
We find that there is no significant difference in the inner fractions of Type Ia and CC SNe
in the Sbc bin ($F_{\rm Ia}=0.32_{-0.06}^{+0.07}$ and $F_{\rm CC}=0.31_{-0.05}^{+0.05}$).
Even selecting unbarred Sb--Sbc bins where the bulge should be (slightly) more prominent,
we get $F_{\rm Ia}=0.31_{-0.05}^{+0.06}$ and $F_{\rm CC}=0.33_{-0.04}^{+0.04}$.
Here, we do not use the earlier morphological bins because they are rarely populated by CC SNe
making it impossible to estimate the $F_{\rm CC}$ (see Table~\ref{table_SN_morph}).
On the other hand, the later morphological bins are not suitable because of the weaker bulge component
\cite[e.g.][]{2009ApJ...705..245O,2011A&A...532A..75D}.

Interestingly, $F_{\rm Ia}=0.32_{-0.05}^{+0.06}$ for 28 Type Ia SNe in all S0--S0/a galaxies,
which is the same as $F_{\rm Ia}=0.32_{-0.02}^{+0.02}$ for 152 SNe Ia in all Sa--Sm hosts
and even the same as $F_{\rm Ia}$ for only unbarred Sbc galaxies.
With the results in Section~\ref{resdiscus_sub1}, this suggests that the deviation of
the radial distribution of Type Ia SNe in S0--S0/a galaxies from that in Sa--Sm hosts
(upper panel of Fig.~\ref{hist_distr} and Table~\ref{diffSNe_KS_AD})
is mostly attributed to the contribution by the outer bulge SNe Ia in S0--S0/a galaxies
(see also the corresponding mean values of $\tilde{r}$ in Fig.~\ref{hist_distr}).

These results confirm that the old bulges of Sa--Sm galaxies
are not significant producers of Type Ia SNe, while
the bulge populations are significant for SNe Ia only in S0--S0/a galaxies.
In S0--S0/a hosts, the relative fraction of bulge to disc SNe Ia is probably changed
in comparison with that in Sa--Sm hosts, because the progenitor population from
the discs of S0--S0/a galaxies should be much lower due to the lower number of
young/intermediate stellar populations.

\begin{table*}
  \centering
  \begin{minipage}{121mm}
  \caption{Consistency of CC and Type Ia SN distributions with an exponential surface
           density model.}
  \tabcolsep 4pt
  \label{tableallSNe}
  \begin{tabular}{llrccccrccc}
  \hline
    \multicolumn{1}{c}{}&&\multicolumn{4}{c}{$\tilde{r}\in[0; \infty)$}&\multicolumn{1}{c}{}&\multicolumn{4}{c}{$\tilde{r}\in[0.2; \infty)$}\\
    \cline{3-6} \cline{8-11}
    \multicolumn{1}{l}{Host}&\multicolumn{1}{c}{SN}&\multicolumn{1}{c}{$N_{\rm SN}$}&\multicolumn{1}{c}{$\langle\tilde{r}\rangle \pm \sigma$}&\multicolumn{1}{c}{$P_{\rm KS}$}&\multicolumn{1}{c}{$P_{\rm AD}$}&&\multicolumn{1}{c}{$N_{\rm SN}$}&\multicolumn{1}{c}{$P_{\rm KS}$}&\multicolumn{1}{c}{$P_{\rm AD}$}&\multicolumn{1}{c}{\scriptsize $\tilde{h}_{\rm SN}$}\\
    \multicolumn{1}{l}{\, (1)}&\multicolumn{1}{c}{(2)}&\multicolumn{1}{c}{(3)}&
    \multicolumn{1}{c}{(4)}&\multicolumn{1}{c}{(5)}&\multicolumn{1}{c}{(6)}&&\multicolumn{1}{c}{(7)}&\multicolumn{1}{c}{(8)}&
    \multicolumn{1}{c}{(9)}&\multicolumn{1}{c}{(10)}\\
  \hline
    S0--S0/a&Ia&28&$0.57\pm0.38$&0.593&0.344&&21&0.297&0.455&$0.32\pm0.04$\\
  \hline
    Sa--Sm&Ia&152&$0.46\pm0.28$&0.209&0.091&&127&0.533&0.488&$0.21\pm0.02$\\
    Sa--Sm (barred)&Ia&73&$0.47\pm0.23$&0.095&0.076&&64&0.421&0.244&$0.21\pm0.03$\\
    Sa--Sm (unbarred)&Ia&79&$0.46\pm0.32$&0.520&0.548&&63&0.962&0.912&$0.22\pm0.02$\\
    Sa--Sbc&Ia&79&$0.46\pm0.29$&0.441&0.295&&69&0.968&0.980&$0.21\pm0.02$\\
    Sa--Sbc (barred)&Ia&39&$0.46\pm0.23$&0.315&0.210&&35&0.595&0.601&$0.20\pm0.03$\\
    Sa--Sbc (unbarred)&Ia&40&$0.47\pm0.34$&0.864&0.874&&34&0.683&0.917&$0.21\pm0.03$\\
    Sc--Sm&Ia&73&$0.46\pm0.27$&0.443&0.265&&58&0.432&0.408&$0.22\pm0.02$\\
    Sc--Sm (barred)&Ia&34&$0.47\pm0.24$&0.279&0.359&&29&0.424&0.477&$0.22\pm0.03$\\
    Sc--Sm (unbarred)&Ia&39&$0.45\pm0.30$&0.666&0.628&&29&0.932&0.792&$0.23\pm0.03$\\
  \hline
    Sa--Sm&CC&319&$0.49\pm0.26$&\textbf{0.001}&\textbf{0.000}&&286&0.075&0.056&$0.21\pm0.01$\\
    Sa--Sm (barred)&CC&136&$0.52\pm0.26$&\textbf{0.045}&\textbf{0.005}&&125&0.062&0.116&$0.23\pm0.02$\\
    Sa--Sm (unbarred)&CC&183&$0.46\pm0.25$&\textbf{0.028}&\textbf{0.008}&&161&0.454&0.359&$0.20\pm0.01$\\
    Sa--Sbc&CC&106&$0.45\pm0.21$&\textbf{0.004}&\textbf{0.008}&&95&0.102&0.118&$0.19\pm0.02$\\
    Sa--Sbc (barred)&CC&50&$0.49\pm0.21$&\textbf{0.050}&\textbf{0.029}&&45&\textbf{0.030}&\textbf{0.046}&$0.21\pm0.04$\\
    Sa--Sbc (unbarred)&CC&56&$0.42\pm0.22$&\textbf{0.035}&0.052&&50&0.384&0.432&$0.17\pm0.03$\\
    Sc--Sm&CC&213&$0.50\pm0.27$&\textbf{0.036}&\textbf{0.004}&&191&0.125&0.190&$0.23\pm0.02$\\
    Sc--Sm (barred)&CC&86&$0.54\pm0.29$&0.118&0.066&&80&0.715&0.785&$0.24\pm0.03$\\
    Sc--Sm (unbarred)&CC&127&$0.48\pm0.26$&0.146&\textbf{0.041}&&111&0.164&0.226&$0.22\pm0.02$\\
  \hline \\
  \end{tabular}
  \parbox{\hsize}{
    Columns~1 and 2 give the subsample;
    Col.~3 is the number of SNe in the subsample for the full radial range $\tilde{r}\in[0; \infty)$;
    Col.~4 is the mean of $\tilde{r}$ with standard deviation;
    Cols.~5 and 6 are the $P_{\rm KS}$ and $P_{\rm AD}$ probabilities from one-sample KS and AD tests,
    respectively, that the distribution of SNe is drawn from the best-fitting exponential surface density profile;
    Cols.~7, 8, and 9 are, respectively, the same as Cols.~3, 5, and 6, but for $\tilde{r}\in[0.2; \infty)$;
    Col.~10 is the maximum likelihood value of $\tilde h_{\rm SN} = h_{\rm SN}/R_{25}$
    with bootstrapped error (repeated $10^4$ times) for the inner-truncated disc.
    The $P_{\rm KS}$ and $P_{\rm AD}$ are calculated using the calibrations by
    \citet{Massey51} and \citet{1986gft..book.....D}, respectively.
    The statistically significant deviations from an exponential law are highlighted in bold.}
  \end{minipage}
\end{table*}

Moreover, we do not detect the relative deficiency of
Type Ia SNe in comparison with CC SNe in the inner regions of spiral hosts,
contrary to \citet[][]{1997AJ....113..197V},
\citet[][]{1997ApJ...483L..29W}, and \citet[][]{2015MNRAS.448..732A}.
Instead, the radial distributions of both types of SNe are well matched between each other in
all the subsamples of Sa--Sbc and Sc--Sm galaxies supporting the idea that the relative
concentration of CC SNe in the centres of spirals found by these authors
is most probably due to a contribution of the central excess of CC SNe in disturbed galaxies
(e.g. \citealt*{2012MNRAS.424.2841H}; Paper~\citetalias{2014MNRAS.444.2428H}),
which are excluded from our sample (see Section~\ref{sample}).

\citet[][]{1997ApJ...483L..29W} discussed the possibilities
that the relative deficiency of Type Ia SNe that they found (and which we do
not confirm) in the inner regions of spiral galaxies may be due to a stronger
dust extinction for Type Ia events than for CC SNe.
They suggested that massive progenitors of CC SNe within associations
might create large cavities in the discs through their own stellar
winds or earlier SN explosions, therefore making the discovery of CC SNe easier.
Despite the higher luminosity of Type Ia SNe at maximum,
their lower mass progenitors make such cavities less likely
\citep[see also][]{1990ApJ...359..277V}.
However, our results discussed above show that dust
extinction in the discs of nearby non-disturbed spirals
should not be different for Type Ia and CC SNe.
Moreover, in Aramyan et al. (in preparation), we show that although CC SNe are
more concentrated to the brightness peaks of spiral arms than are Type Ia events,
both SN types occur mostly in spiral arms
\citep*[see also][]{1994PASP..106.1276B,1996ApJ...473..707M}
where large cavities, if present,
should be shared, on average, between the progenitors of both SN types.

The differences between our results and those of previous studies is that
we deproject and normalize the galactocentric radii of SNe,
while the other studies used projected and normalized \citep[][]{1997AJ....113..197V},
or projected linear \citep[][]{1997ApJ...483L..29W},
or normalized to flux \citep[][]{2015MNRAS.448..732A} galactocentric radii.
Another important difference is that, contrary to our study, \citet{1997ApJ...483L..29W}
and \citet{2015MNRAS.448..732A} included highly inclined galaxies in their studies.

\subsection{The deprojected exponential surface density distribution}
\label{resdiscus_sub3}

\begin{table*}
  \centering
  \begin{minipage}{121mm}
  \caption{Consistency of Types Ibc and II SN distributions with an
           exponential surface density model.}
  \tabcolsep 4pt
  \label{tableallCC}
  \begin{tabular}{llrccccrccc}
  \hline
    \multicolumn{1}{c}{}&&\multicolumn{4}{c}{$\tilde{r}\in[0; \infty)$}&\multicolumn{1}{c}{}&\multicolumn{4}{c}{$\tilde{r}\in[0.2; \infty)$}\\
    \cline{3-6} \cline{8-11}
    \multicolumn{1}{l}{Host}&\multicolumn{1}{c}{SN}&\multicolumn{1}{c}{$N_{\rm SN}$}&\multicolumn{1}{c}{$\langle\tilde{r}\rangle \pm \sigma$}&\multicolumn{1}{c}{$P_{\rm KS}$}&\multicolumn{1}{c}{$P_{\rm AD}$}&&\multicolumn{1}{c}{$N_{\rm SN}$}&\multicolumn{1}{c}{$P_{\rm KS}$}&\multicolumn{1}{c}{$P_{\rm AD}$}&\multicolumn{1}{c}{\scriptsize $\tilde{h}_{\rm SN}$}\\
    \multicolumn{1}{l}{\, (1)}&\multicolumn{1}{c}{(2)}&\multicolumn{1}{c}{(3)}&
    \multicolumn{1}{c}{(4)}&\multicolumn{1}{c}{(5)}&\multicolumn{1}{c}{(6)}&&\multicolumn{1}{c}{(7)}&\multicolumn{1}{c}{(8)}&
    \multicolumn{1}{c}{(9)}&\multicolumn{1}{c}{(10)}\\
  \hline
    Sa--Sm&Ibc&80&$0.40\pm0.20$&0.063&\textbf{0.043}&&67&0.240&0.396&$0.17\pm0.02$\\
    Sa--Sm (barred)&Ibc&32&$0.42\pm0.22$&0.485&0.383&&27&0.698&0.649&$0.18\pm0.02$\\
    Sa--Sm (unbarred)&Ibc&48&$0.38\pm0.19$&0.121&0.105&&40&0.430&0.713&$0.16\pm0.02$\\
    Sa--Sbc&Ibc&31&$0.35\pm0.15$&0.112&0.093&&25&0.127&0.160&$0.14\pm0.02$\\
    Sa--Sbc (barred)&Ibc&12&$0.37\pm0.18$&0.308&0.536&&9&0.204&0.251&$0.17\pm0.03$\\
    Sa--Sbc (unbarred)&Ibc&19&$0.34\pm0.12$&0.170&0.114&&16&0.419&0.482&$0.13\pm0.02$\\
    Sc--Sm&Ibc&49&$0.43\pm0.23$&0.267&0.277&&42&0.477&0.855&$0.18\pm0.03$\\
    Sc--Sm (barred)&Ibc&20&$0.45\pm0.24$&0.670&0.591&&18&0.792&0.984&$0.19\pm0.03$\\
    Sc--Sm (unbarred)&Ibc&29&$0.41\pm0.23$&0.589&0.616&&24&0.881&0.911&$0.18\pm0.03$\\
  \hline
    Sa--Sm&II&239&$0.52\pm0.27$&\textbf{0.005}&\textbf{0.000}&&219&0.063&0.052&$0.23\pm0.01$\\
    Sa--Sm (barred)&II&104&$0.55\pm0.27$&0.074&\textbf{0.007}&&98&0.113&0.131&$0.24\pm0.02$\\
    Sa--Sm (unbarred)&II&135&$0.49\pm0.26$&0.093&\textbf{0.024}&&121&0.393&0.323&$0.22\pm0.02$\\
    Sa--Sbc&II&75&$0.49\pm0.23$&\textbf{0.012}&\textbf{0.022}&&70&0.323&0.220&$0.21\pm0.02$\\
    Sa--Sbc (barred)&II&38&$0.52\pm0.20$&\textbf{0.029}&\textbf{0.029}&&36&\textbf{0.022}&\textbf{0.050}&$0.22\pm0.03$\\
    Sa--Sbc (unbarred)&II&37&$0.46\pm0.25$&0.097&0.229&&34&0.588&0.668&$0.19\pm0.03$\\
    Sc--Sm&II&164&$0.53\pm0.28$&0.079&\textbf{0.008}&&149&0.189&0.155&$0.24\pm0.01$\\
    Sc--Sm (barred)&II&66&$0.56\pm0.30$&0.176&0.107&&62&0.571&0.711&$0.25\pm0.03$\\
    Sc--Sm (unbarred)&II&98&$0.50\pm0.27$&0.140&0.051&&87&0.186&0.197&$0.23\pm0.02$\\
  \hline \\
  \end{tabular}
  \parbox{\hsize}{
    The explanation for the columns is the same as for Table~\ref{tableallSNe}.
    In the SNe II subsample, there are 18 Type IIb SNe (see Table~\ref{table_SN_morph})
    with $\langle\tilde{r}\rangle \pm \sigma =0.61\pm0.25$, suggesting that in terms of
    the radial distribution they likely belong to SNe II rather than to SNe Ibc group.
    The statistically significant deviations from an exponential law are highlighted in bold.}
  \end{minipage}
\end{table*}
\begin{figure*}
\begin{center}$
\begin{array}{@{\hspace{0mm}}l@{\hspace{0mm}}r@{\hspace{0mm}}}
\includegraphics[width=0.5\hsize]{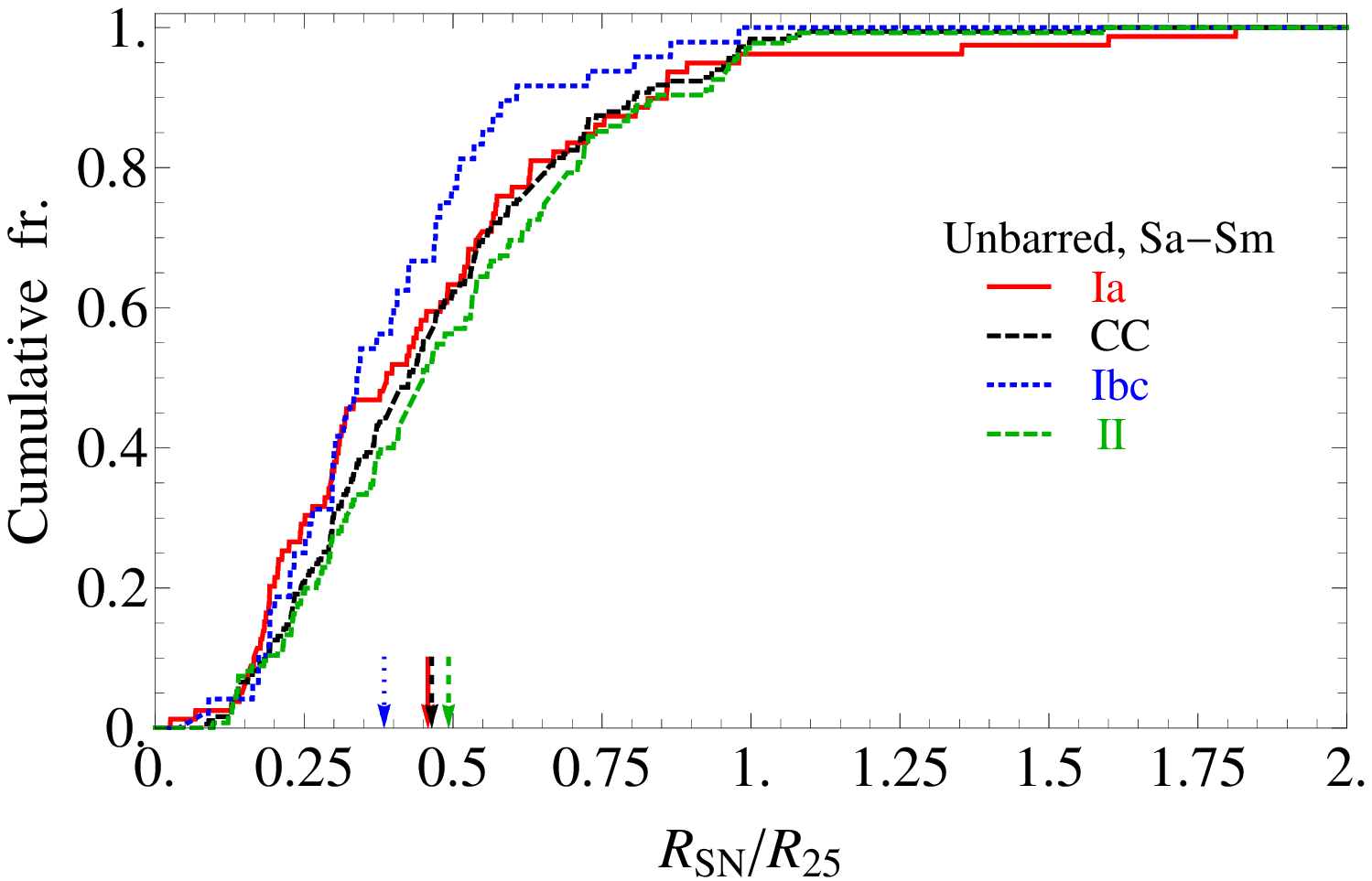} &
\includegraphics[width=0.5\hsize]{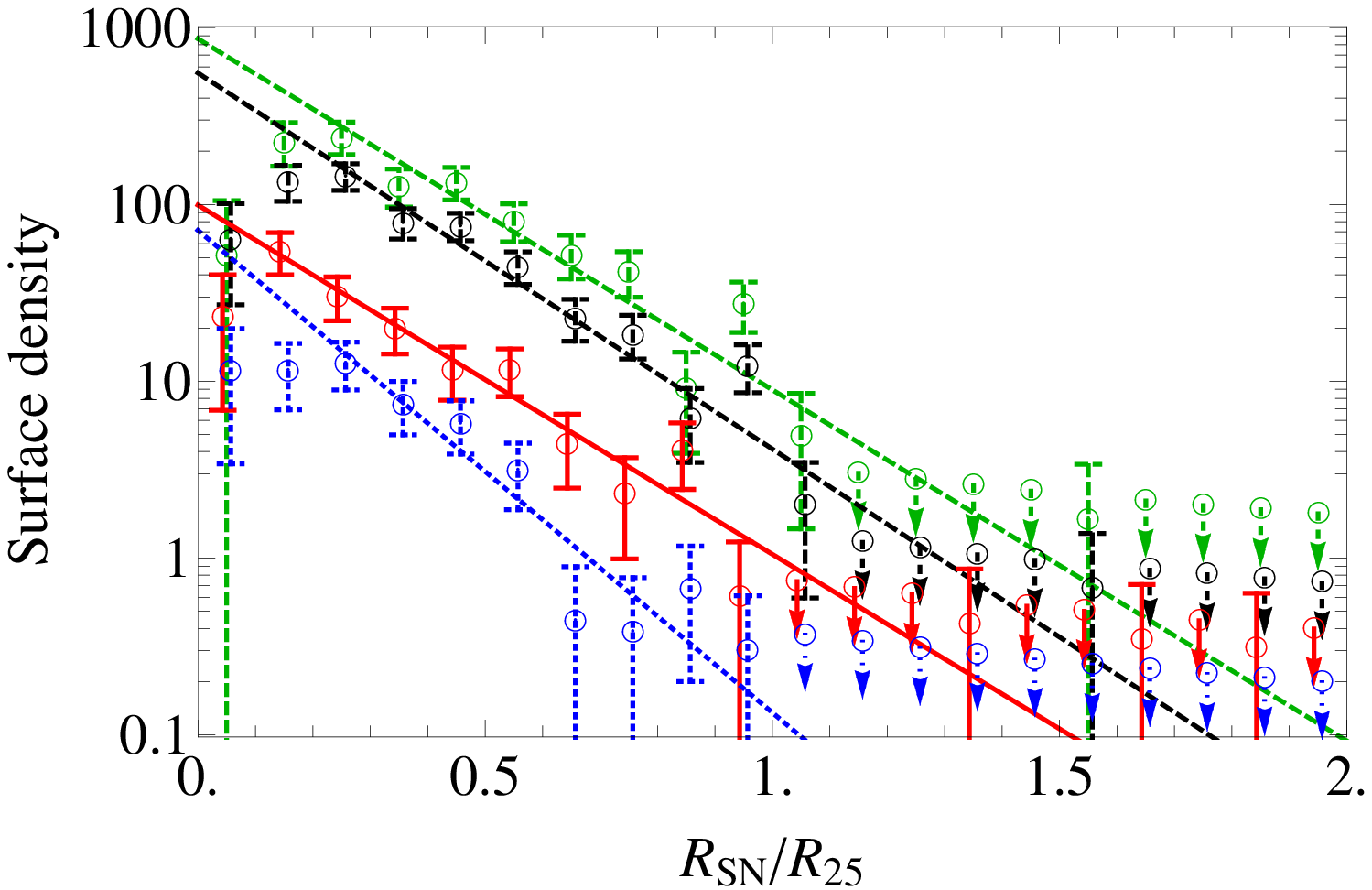}
\end{array}$
\end{center}
\caption{Left: cumulative fractions of the different types of SNe versus deprojected and
         normalized galactocentric radius ($\tilde{r}=R_{\rm SN}/R_{25}$)
         in unbarred Sa--Sm host galaxies. The mean $\tilde{r}$ values of
         each SNe type are shown by arrows.
         Right: surface density distributions (with arbitrary normalization) of the different types of
         SNe in the same subsample of hosts.
         The different lines show the maximum likelihood exponential surface density profiles
         estimated for the inner-truncated disc.
         The error bars assume a Poisson distribution.
         Down arrows represent the upper-limits of surface density (with $\pm1$ SN if none
         is found).
         For better visibility, the distributions are shifted vertically sorted by
         increasing the mean $\tilde{r}$ as one moves upwards.}
\label{cum_SDD}
\end{figure*}

It is widely accepted that the surface density
distribution of SNe in discs follows an exponential law
(e.g. \citealt{1975A&A....44..267B,1977MNRAS.178..693V,1992A&A...264..428B,
1997AJ....113..197V}; \citetalias{2009A&A...508.1259H};
\citealt{2010MNRAS.405.2529W,2013Ap&SS.347..365N}).
However, a comprehensive analysis of the surface density distribution
in different samples of barred and unbarred galaxies has not been performed
and this is one of the main goals of the present study.

Following \citetalias{2009A&A...508.1259H},
we fit an exponential surface density profile, $\Sigma(R)$, to the
distribution of deprojected normalized radii, using maximum likelihood
estimation (MLE). Here,
$\Sigma^{\rm SN}(\tilde{r})=\Sigma_0^{\rm SN} \exp(-\tilde{r}/\tilde{h}_{\rm SN})$,
where $\tilde{h}_{\rm SN}$ is the scale length of the distribution
and $\Sigma_0^{\rm SN}$ is the central surface density of SNe.
To check whether the distribution of SN radii follows the best-fitting
exponential surface density profile, we perform one-sample KS and AD tests
on the normalized cumulative distributions of SNe, where the exponential model has a
cumulative normalized distribution
$E(\tilde{r}) = 1 - (1 + {\tilde{r}/\tilde{h}_{\rm SN}}) \exp (- {\tilde{r}/ \tilde{h}_{\rm SN}})$.

In columns~3--6 of Table~\ref{tableallSNe}, the total number of SNe in
the full radial range, their mean radius with standard deviation,
the KS and AD test $P$-values,
$P_{\rm KS}$ and $P_{\rm AD}$ are, respectively, presented for the different subsamples.
In Table~\ref{tableallCC}, despite the small number statistics of
Type Ibc SNe (see Table~\ref{table_SN_morph}), we consider Types Ibc and II SNe separately.

From columns~5 and 6 of Tables~\ref{tableallSNe} and \ref{tableallCC},
we see that in many subsamples of CC SNe, in contrast to Type Ia SNe, the surface density distribution
is not consistent with an exponential profile.
Fig.~\ref{hist_distr} hints that the observed
inconsistency is probably due to
the central $\tilde{r} \lesssim 0.2$ deficit of SNe
(farther, see also in the right-hand panel of Fig.~\ref{cum_SDD}).
For this reason, we repeat the above described tests for $\tilde{r}\in[0.2; \infty)$ range
and find that the inconsistency vanishes in most of the subsamples
(see columns~8 and 9 in Tables~\ref{tableallSNe} and \ref{tableallCC}).
The corresponding scale lengths for the inner-truncated disc are listed
in column~10 of Tables~\ref{tableallSNe} and \ref{tableallCC}.
As expected, only the distribution of CC SNe in early-type barred spirals
is inconsistent with an exponential distribution due to the impact of bars on
the radial distribution of CC SNe as discussed in Section~\ref{resdiscus_sub2}.
The effect is not seen for Type Ibc SNe probably due to
the small number statistics of these SNe in early-type barred spirals
(see column~7 in Tables~\ref{tableallCC}).

Finally, to eliminate the effects induced by bars, we compare the radial distributions of
Types Ibc and II SNe in unbarred spiral galaxies only.
The two-sample KS and AD tests show that the radial distribution of Type Ibc SNe is highly inconsistent
with that of Type II ($P_{\rm KS}=0.026$ and $P_{\rm AD}=0.014$),
because the former are more centrally concentrated
(see the mean of $\tilde{r}$ and $\tilde h_{\rm SN}$ values in Table~\ref{tableallCC}).

The radial position within host galaxies can be used as a proxy for the local metallicity
since the short lifetime of CC SN progenitor (tens of Myr)
is not enough to allow far migration from its birthplace.
Therefore, the physical explanation for the more concentrated distribution of SNe Ibc
with respect to SNe II in non-disturbed and unbarred spiral galaxies is that SNe Ibc
arise from more metal-rich environments,\footnote{{\footnotesize
The top-heavy initial mass function and/or enhanced close binary fraction
in the central regions of strongly disturbed/interacting galaxies might play
an important role in explaining the inner excess of SNe Ibc compared to SNe II
\cite[e.g.][]{2012MNRAS.424.2841H,2013MNRAS.436.3464K,2015PASA...32...19A}.
However, the roles of these factors are difficult to estimate,
since our analysis is restricted to non-disturbed host galaxies.}}
as has been widely discussed
(e.g. \citealt{1997AJ....113..197V,2005AJ....129.1369P,2008Ap.....51...69H};
\citealt*{2008ApJ...673..999P};
\citealt{2009A&A...503..137B,2009MNRAS.399..559A};
\citetalias{2009A&A...508.1259H};
\citealt{2012MNRAS.424.1372A}).
Here, we do not go into deeper considerations, instead we refer the reader to
the mentioned references for more complete discussions.

Despite the smaller numbers statistics, we analyse the radial distributions
of Types Ib and Ic SNe in unbarred galaxies
(18 SNe Ib and 23 SNe Ic; see Table~\ref{table_SN_morph}).
Compared with the SNe Ibc versus SNe II test, the radial distributions of
SNe Ib and SNe Ic are sufficiently similar that the two-sample KS and AD tests
fail to distinguish them with statistical significance
($P_{\rm KS}=0.119$ and $P_{\rm AD}=0.202$).
In the inner-truncated disc,
the scale length of Type Ib SNe ($0.14\pm0.03$) is not significantly lower
from that of Type Ic SNe ($0.17\pm0.03$),
while the scale length of all the Ibc family is between these values
(see Table~\ref{tableallCC}).

Fig.~\ref{cum_SDD} presents the cumulative and
surface density distributions of the different types of SNe
in unbarred Sa--Sm host galaxies.
According to the scale lengths in Tables~\ref{tableallSNe} and \ref{tableallCC},
the fit lines in Fig.~\ref{cum_SDD} show the exponential surface density profiles of
SNe in the same subsample.
A central ($\tilde{r} \lesssim 0.2$) drop from an exponential distribution
is observed for all the SNe types, less prominent for SNe Ia.

It is important to note that the scale lengths of SNe ($\tilde h_{\rm SN} = h_{\rm SN}/R_{25}$) in
Tables~\ref{tableallSNe} and \ref{tableallCC} are calculated using
the $\mu_g=25\,\rm mag\,arcsec^{-2}$ isophotal level in the SDSS $g$-band (see Section~\ref{sample}),
while the previous papers used the same magnitude isophotal level in $B$-band
(e.g. \citealt{2004AstL...30..729T};
\citetalias{2009A&A...508.1259H};
\citealt{2010MNRAS.405.2529W,2012A&A...540L...5H,2013MNRAS.436.3464K}).
In Paper~\citetalias{2012A&A...544A..81H}, we showed that galaxies have sizes
systematically larger in the $g$-band than in the $B$-band
(see fig.~11 and table~4 in Paper~\citetalias{2012A&A...544A..81H})
due to the fact that our $g$-band measurements are performed at the equivalent
of the $\left\langle\mu_B\right\rangle\simeq 25.48$ isophote,
hence at typically lower surface brightness threshold.
Therefore, considering that our radii of host galaxies are greater than those in the HyperLeda
on average by a factor of $1.32\pm0.01$ (see Paper~\citetalias{2012A&A...544A..81H}),
we get generally about 25 per cent smaller scale lengths of SNe in $g$-band compared
with the earlier estimations in $B$-band (e.g. see \citetalias{2009A&A...508.1259H} for CC SNe).

\section{Conclusions}
\label{concl}

In this third paper of a series, using a well-defined and homogeneous
sample of SNe and their host galaxies from the coverage of SDSS DR10, we analyse the impact
of bars and bulges on the radial distributions of the different types of SNe
in the stellar discs of host galaxies with various morphologies.
Our sample consists of 419 nearby (${\leq {\rm 100~Mpc}}$),
low-inclination ($i \leq 60^\circ$), and morphologically non-disturbed S0--Sm galaxies,
hosting 500 SNe in total.

All the results that we summarize below concerning the radial distributions of SNe
in barred galaxies can be explained considering the strong impact of the bars on
the distribution of massive star formation in stellar discs of galaxies,
particularly in early-type spirals.
On the other hand, the bulge component of Type Ia SNe distribution shows
a negligible impact on the radial distribution of SNe Ia,
except in S0-S0/a galaxies.

We also check that there are no strong selection effects and biases within our
SNe and host galaxies samples, which could drive the observed behaviours of
the radial distributions of Type Ia and CC SNe in the disc galaxies presented in this study.

The results obtained in this article are summarized below,
along with their interpretations.

\begin{enumerate}
\item In Sa--Sm galaxies, all CC and the vast majority of Type Ia SNe belong to the disc,
      rather than the bulge component (Fig.~\ref{hist_coord} and Table~\ref{SNeindisc}).
      The result suggests that the rate of SNe Ia in spiral galaxies is dominated by
      a relatively young/intermediate progenitor population.
      This observational fact makes the deprojection of galactocentric radii of both
      types of SNe a key point in the statistical studies of their distributions.
\item The radial distribution of Type Ia SNe in S0--S0/a galaxies is inconsistent
      with that in Sa--Sm hosts (as seen in Fig.~\ref{hist_distr} and
      Table~\ref{diffSNe_KS_AD} for the AD statistic but only
      very marginally so in the KS statistic).
      This inconsistency is mostly attributed to the contribution
      by SNe Ia in the outer bulges of S0--S0/a galaxies.
      In these hosts, the relative fraction of bulge to disc SNe Ia is probably changed
      in comparison with that in Sa--Sm hosts, because the progenitor population from
      the discs of S0--S0/a galaxies should be much lower due to the lower number of
      young/intermediate stellar populations.
\item The radial distribution of CC SNe in barred Sa--Sbc galaxies is not consistent
      with that of unbarred Sa--Sbc hosts (Fig.~\ref{hist_distr} and Table~\ref{diffSNe_KS_AD}),
      while for Type Ia SNe the distributions are
      not significantly different (Table~\ref{diffSNe_KS_AD}).
      At the same time, the radial distributions of both
      Type Ia and CC SNe in Sc--Sm galaxies are not affected by bars
      (Table~\ref{diffSNe_KS_AD}).
      These results are explained by a substantial suppression of
      massive star formation in the radial range swept by strong bars of
      early-type barred galaxies.
\item The radial distribution of CC SNe in unbarred Sa--Sbc galaxies is more centrally
      peaked and inconsistent with that in unbarred Sc--Sm hosts
      (as seen in Fig.~\ref{hist_distr} and Table~\ref{diffSNe_KS_AD} for the KS statistic but only
      marginally so in the AD statistic). On the other hand, the radial distribution
      of Type Ia SNe in unbarred galaxies is not affected by host morphology (Table~\ref{diffSNe_KS_AD}).
      These results can be well explained by the distinct distributions of massive stars
      in the discs of early- and late-type spirals.
      In unbarred Sa--Sbc galaxies, star formation is more concentrated to the inner regions
      (H$\alpha$ emission outside the nucleus) and should thus be responsible
      for the observed radial distribution of CC SNe.
\item The radial distribution of CC SNe, in contrast to Type Ia SNe, is inconsistent with
      the exponential surface density profile (Tables~\ref{tableallSNe} and
      \ref{tableallCC}), because of the central ($\tilde{r} \lesssim 0.2$) deficit of SNe.
      However, in the $\tilde{r}\in[0.2; \infty)$ range, the inconsistency
      vanishes for CC SNe in most of the subsamples of spiral galaxies.
      In the inner-truncated disc, only the radial distribution of CC SNe in barred early-type spirals
      is inconsistent with an exponential surface density profile, which
      appears to be caused by the impact of bars on the radial distribution of CC SNe.
\item In the inner regions of non-disturbed spiral hosts, we do not detect
      a relative deficiency of Type Ia SNe
      with respect to CC (Table~\ref{table0.3ratios}), contrary to what was found by other authors, who
      had explained this by possibly stronger dust extinction for Type Ia than for CC SNe.
      Instead, the radial distributions of both types of SNe are similar
      in all the subsamples of Sa--Sbc and Sc--Sm galaxies (Table~\ref{diffSNe_KS_AD}), which supports
      the idea that the relative increase of CC SNe in the inner regions of spirals found by the other
      authors is most probably due to the central excess of CC SNe in disturbed galaxies.
\item As was found in earlier studies, the physical explanation for the more concentrated distribution of
      SNe Ibc with respect to SNe II in non-disturbed and unbarred spiral galaxies
      (Fig.~\ref{cum_SDD}) is that SNe Ibc arise from more metal-rich environments.
      The radial distributions of Types Ib and Ic SNe
      are sufficiently similar that the KS and AD tests fail to distinguish
      them with statistical significance.
\end{enumerate}

\section*{Acknowledgements}

We would like to thank the anonymous referee for his/her excellent commentary,
and also V.~de~Lapparent, E.~Bertin, and T.A.~Nazaryan for their
constructive comments on the earlier drafts of this manuscript.
AAH, AGK, LVB, and ARP acknowledge the hospitality of the
Institut d'Astrophysique de Paris (France) during their
stay as visiting scientists supported by
the Programme Visiteurs Ext\'{e}rieurs (PVE).
This work was supported by State Committee Science MES RA,
in frame of the research project number SCS~13--1C013.
AAH is also partially supported by the ICTP.
VA is supported by grant SFRH/BPD/70574/2010 from FCT (Portugal).
DK acknowledges financial support from the Centre
National d'\'{E}tudes Spatiales (CNES).
MT is partially supported by the PRIN-INAF 2011 with the project
Transient Universe: from ESO Large to PESSTO.
This work was made possible in part by a research grant from the
Armenian National Science and Education Fund (ANSEF)
based in New York, USA.
Funding for SDSS-III has been provided by the Alfred P.~Sloan Foundation,
the Participating Institutions, the National Science Foundation,
and the US Department of Energy Office of Science.
The SDSS--III web site is \href{http://www.sdss3.org/}{http://www.sdss3.org/}.
SDSS--III is managed by the Astrophysical Research Consortium for the
Participating Institutions of the SDSS--III Collaboration including the
University of Arizona, the Brazilian Participation Group,
Brookhaven National Laboratory, University of Cambridge,
University of Florida, the French Participation Group,
the German Participation Group, the Instituto de Astrofisica de Canarias,
the Michigan State/Notre Dame/JINA Participation Group,
Johns Hopkins University, Lawrence Berkeley National Laboratory,
Max Planck Institute for Astrophysics, New Mexico State University,
New York University, Ohio State University, Pennsylvania State University,
University of Portsmouth, Princeton University, the Spanish Participation Group,
University of Tokyo, University of Utah, Vanderbilt University,
University of Virginia, University of Washington, and Yale University.

\bibliography{snbibIII}

\section*{Supporting information}

Additional Supporting Information may be found in
the online version of this article:\\
\\
\textbf{PaperIIIonlinedata.csv}
\\
(\href{http://www.mnras.oxfordjournals.org/lookup/suppl/doi:10.1093/mnras/stv2853/-/DC1}
{http://www.mnras.oxfordjournals.org/lookup/suppl/doi:10.1093/}
\href{http://www.mnras.oxfordjournals.org/lookup/suppl/doi:10.1093/mnras/stv2853/-/DC1}
{mnras/stv2853/-/DC1}).
\\
\\
Please note: Oxford University Press is not responsible for the
content or functionality of any supporting materials supplied by
the authors. Any queries (other than missing material) should be
directed to the corresponding author for the article.

\label{lastpage}

\end{document}